\documentclass[twoside,english]{article}

\usepackage[latin1]{inputenc}
\usepackage{geometry}
\usepackage{graphicx}
\usepackage{amssymb}

\makeatletter

 \newcommand{\lyxaddress}[1]{
   \par {\raggedright #1 
   \vspace{1.4em}
   \noindent\par}
 }

\usepackage{babel}
\makeatother
\begin{document}

\title{\textbf{Coherence oscillations in dephasing by non-Gaussian shot
noise}}

\author{Izhar Neder$^{\textrm{1}}$ and Florian Marquardt$^{\textrm{2}}$}

\maketitle

\lyxaddress{(1) Braun Center for Submicron Research, Department of Condensed
Matter Physics, Weizmann Institute of Science, Rehovot 76100, Israel\\
(2) Physics Department, Center for NanoScience, and Arnold Sommerfeld
Center for Theoretical Physics, Ludwig-Maximilians Universit\"at
M\"unchen, Theresienstr. 37, 80333 Munich, Germany}

\begin{abstract}
A non-perturbative treatment is developed for the dephasing produced
by the shot noise of a one-dimensional electron channel. It is applied
to two systems: a charge qubit and the electronic Mach-Zehnder interferometer,
both of them interacting with an adjacent partitioned electronic channel
acting as a detector. We find that the visibility (interference contrast)
can display oscillations as a function of detector voltage and interaction
time. This is a unique consequence of the non-Gaussian properties
of the shot noise, and only occurs in the strong coupling regime,
when the phase contributed by a single electron exceeds $\pi$. The
resulting formula reproduces the recent surprising experimental observations
reported in {[}I. Neder et al., cond-mat/0610634{]}, and indicates
a general explanation for similar visibility oscillations observed
earlier in the Mach-Zehnder interferometer at large bias voltage.
We explore in detail the full pattern of oscillations as a function
of coupling strength, voltage and time, which might be observable
in future experiments.
\end{abstract}

\section{Introduction}

\newcommand{\pra}{Phys. Rev. A}
\newcommand{\prb}{Phys. Rev. B}
\newcommand{\pre}{Phys. Rev. E}
\newcommand{\prl}{Phys. Rev. Lett.}

Decoherence, i.e. the destruction of quantum mechanical interference
effects, is a topic whose importance ranges from more fundamental
questions like the quantum-classical crossover to possible applications
of quantum coherent phenomena, such as sensitive measurements and
quantum information and quantum computing. In mesoscopic transport
experiments, decoherence (also called dephasing) is responsible for
the nontrivial temperature- and voltage-dependence of the electrical
conductance in disordered samples (displaying weak localization and
universal conductance fluctuations) and solid-state electron interferometers.

The most important paradigmatic quantum-dissipative models ({}``Caldeira-Leggett''
\cite{1981_01_CaldeiraLeggett_TunnelingWithDissipation,1983_CaldeiraLeggett_QuantumBrownianMotion}
and {}``spin-boson'' \cite{1987_Leggett_ReviewSpinBoson,2000_Weiss_QuantumDissipativeSystems})
and many well-known techniques used for describing decoherence assume
the environment to be a bath of harmonic oscillators, where the fluctuations
obey Gaussian statistics. This assumption is correct for some cases
(e.g. photons and phonons), and generally represents a very good approximation
for the combined contribution of many weakly coupled fluctuators,
due to the central limit theorem. However, ultrasmall structures may
couple only to a few fluctuators (spins, charged defects etc.), thus
requiring models of dephasing by \emph{non-Gaussian} noise. Such models
are becoming very important right now in the context of quantum information
processing \cite{2002_06_PaladinoFazio_TelegraphNoisePRL,2002_Gassmann_NonlinearBath,2004_04_MakhlinShnirman_DephasingOptimalPoints,2004_08_Martinis_JunctionResonators,2004_12_Astafiev_StrangeNoiseSpectrum,2005_08_Lerner_QuantumTelegraphNoiseLongTime,2005_10_EsteveShnirman_DecoherenceSupraQubit,2005_12_Souse_TrappingCenter,2006_01_SchrieflShnirman_DecoherenceTLS,2006_03_GalperinAltshuler_NonGaussianQubitDecoherence}.
\begin{figure}
\begin{center}\includegraphics[%
  width=0.9\columnwidth]{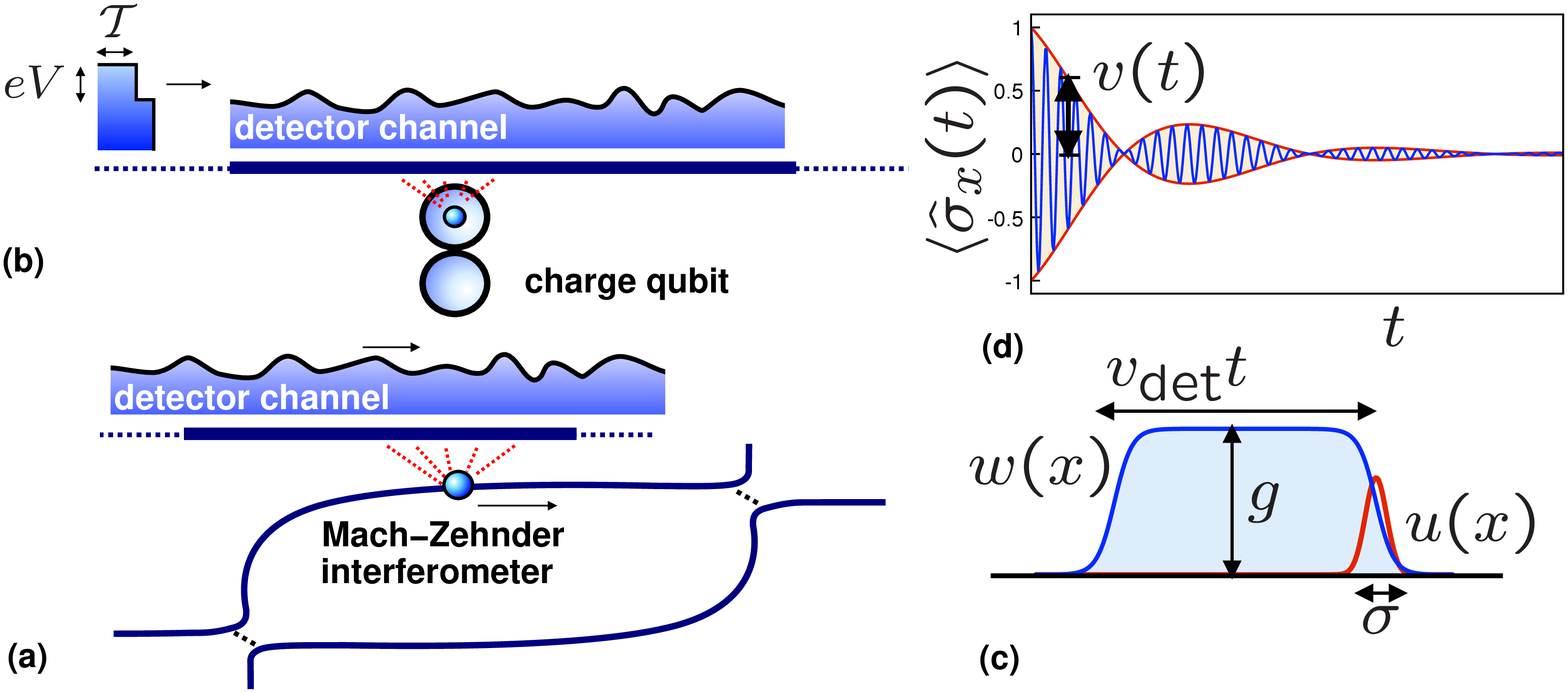}\end{center}

\caption{\label{fig0}Schematic drawing of the models considered in the text:
A detector channel with shot noise coupled to (\textbf{a}) an interferometer
channel or (\textbf{b}) a charge qubit. (\textbf{c}) The interaction
potential $u(x)$ defines the phase function $w(x)$, whose height
gives the dimensionless coupling constant $g$, see equation (\ref{wx}).
(\textbf{d}) Sketch of the time-evolution of the oscillations in $\left\langle \hat{\sigma}_{x}(t)\right\rangle $,
indicating the visibility $v(t)$ as the magnitude of the oscillation
envelope. In this schematic example, the visibility itself oscillates
- this is impossible in models of Gaussian noise but a direct feature
of the non-Gaussian nature of the shot noise, in the strong coupling
regime $g>\pi$ (see text and following figures). }
\end{figure}

Moreover, the quantum measurement process itself is accompanied by
unavoidable fluctuations which dephase the quantum system \cite{1992_BraginskyKhalili_QuantumMeasurement,2003_04_Clerk_QLimitsMesoscopicDetectors,2004_12_Gavish_FermionicAmplifier},
while dephasing itself can conversely be viewed as a kind of detection
process \cite{1990_04_SAI,2002_Imry_MesoscopicPhysics}. Therefore,
\char`\"{}controlled dephasing\char`\"{} experiments can be used to
study the transition from quantum to classical behavior, e.g. by coupling
an electron interferometer to a tunable {}``which path detector''
\cite{1997_08_Levinson_DephasingByQPC_EPL,1997_11_AleinerWingreenMeir_WhichPath,1998_02_Heiblum_WhichPath,2000_06_SprinzakHeiblum_ControlledDephasingOfDoubleDot,2006_07_NguyenLevinson_QDotDephasingByFQHE,2006_07_Rohrlich_ControlledDephasing},
which produces shot noise by partitioning an electron stream \cite{1987_Khlus_Shotnoise,1989_Lesovik_Shotnoise,1990_Buettiker_ScatteringTheorySN,1996_deJongBeenakker_ReviewSN,2000_BlanterBuettiker_ReviewSN}.
In previous mesoscopic controlled dephasing experiments the coupling
between detector and interferometer was weak, requiring the passage
of many detector electrons in order to determine the path. Under these
conditions, the phase of the interfering electron fluctuates according
to a Gaussian random process. 

Recently, a controlled dephasing experiment was performed \cite{2006_Neder_QuantumEraser,2006_07_MZ_DephasingNonGaussianNoise_NederMarquardt}
using an electronic Mach-Zehnder Interferometer (MZI) \cite{2003_Heiblum_MachZehnder,2006_01_Neder_VisibilityOscillations}
coupled to a nearby partitioned edge-channel serving as a detector.
Its results differed substantially from those of earlier controlled-dephasing
experiments. The interference contrast of the Aharonov-Bohm oscillations,
quantified by the visibility $v=(I_{{\rm max}}-I_{{\rm min}})/(I_{{\rm max}}+I_{{\rm min}})$,
revealed two unexpected effects:

(a) The visibility as a function of the detector transmission probability
$\mathcal{T}$ changes from the expected smooth parabolic suppression
$\propto\mathcal{T}(1-\mathcal{T})$ at low detector voltages to a
sharp {}``V-shape'' behaviour at some larger voltages.

(b) The visibility drops to zero at intermediate voltages, then reappears
again as $V$ increases, and vanishes at even larger voltages, thus
displaying \emph{oscillations}.

As estimated in \cite{2006_Neder_QuantumEraser}, three (or even fewer)
detecting electrons suffice to quench the visibility. For this reason,
one suspects that these effects may be a signature of the strong coupling
between interferometer and detector. Indeed, that coupling has already
been exploited to entangle the interfering electrons with the detector
electrons, and afterwards recover the phase information by cross-correlating
the current fluctuations of the MZI and the detector \cite{2006_Neder_QuantumEraser},
even after it has completely vanished in conductance measurements.
The dephasing in the MZI system is caused by the detector's shot noise,
which is known to obey binomial, i.e. non-Gaussian, statistics. Thus,
earlier theoretical discussions of dephasing in the electronic Mach-Zehnder
interferometer, based on a Gaussian environment \cite{2001_SeeligBuettiker_MZDephasing,2004_Marquardt_MZ_PRL,2004_Marquardt_MZ_PRB,2004_10_Marquardt_MZQB_PRL,2005_Foerster_MZ_FCS,2006_04_MZQB_Long},
are no longer sufficient (see \cite{2006_04_LawFeldmanGefen_FQHE_MZ,2006_09_Sukhorukov_MZ_CoupledEdges}
for a discussion of Luttinger liquid physics in an MZI). At the same
time, a nonperturbative treatment is required, to capture the non-Gaussian
effects. Higher moments of the noise become important, and dephasing
starts to depend on the full counting statistics, which itself represents
a topic attracting considerable attention nowadays \cite{1996_06_LevitovLeeLesovik_FullCountingStatistics,2005_Foerster_MZ_FCS,2005_09_AverinSukhorukov_FCS_QPCdetectorPRL}.
The relation between full counting statistics, detection and dephasing
has been explored recently by Averin and Sukhorukov \cite{2005_09_AverinSukhorukov_FCS_QPCdetectorPRL}.
There, the dephasing rate and the measurement rate were considered,
i.e. the focus was placed on the long-time limit, similarly to other
calculations of dephasing by non-Gaussian noise \cite{2002_06_PaladinoFazio_TelegraphNoisePRL,2004_04_MakhlinShnirman_DephasingOptimalPoints,2005_08_Lerner_QuantumTelegraphNoiseLongTime,2005_12_Souse_TrappingCenter,2006_01_SchrieflShnirman_DecoherenceTLS,2006_03_GalperinAltshuler_NonGaussianQubitDecoherence}.
In contrast, we will emphasize the surprising evolution of the visibility
at short to intermediate times. 

The main purpose of this paper then is to present a nonperturbative
treatment of a theoretical model that explains the new experimental
results, and provides quantitative predictions for the behavior of
the visibility as a function of detector bias and partitioning. Furthermore,
we will show how the approximate solution for the MZI is directly
related to an exact solution for the pure dephasing of a charge qubit
by shot noise, where the time evolution of the visibility parallels
the evolution with detector voltage. In conclusion, it will emerge
that the novel features observed in \cite{2006_07_MZ_DephasingNonGaussianNoise_NederMarquardt},
and the results derived here, are in fact fundamental and generic
consequences of dephasing by the non-Gaussian shot noise of a strongly
coupled electron system. As a side effect, this may indicate a solution
to the puzzling observation of visibility oscillations in a MZI without
adjacent detector channel \cite{2006_01_Neder_VisibilityOscillations}.

The paper is organized as follows: We first describe dephasing of
a charge qubit, being the simpler model that can be solved exactly.
After introducing the model in section \ref{sub:Model}, we derive
the exact solution (\ref{sub:Time-evolution-of-the}). We briefly
discuss the relation to full counting statistics (\ref{sub:Relation-to-full}),
and provide formulas obtained in the well-known Gaussian approximation
(\ref{sub:Gaussian-approximation}, \ref{sub:Results-for-the}) for
comparison, before presenting and discussing the results obtained
from a numerical evaluation of the exact expression (section \ref{sub:Exact-numerical-results}).
In section \ref{sec:Nonequilibrium-part-of}, we then go on to introduce
a certain approximation that keeps only the nonequilibrium part of
the noise and allows an analytical discussion of many features, some
of which become particularly transparent in the wave packet picture
of shot noise (\ref{sub:Wave-packet-picture}). In section \ref{sub:Comparison-with-dephasing}
we briefly contrast the features of our solution with those of the
well-known model describing dephasing by classical random telegraph
noise. The Mach-Zehnder interferometer is then described in section
\ref{sec:Electronic-Mach-Zehnder-interferometer}, by first solving
exactly the problem of a single electron interacting with the detector
(\ref{sub:ModelMZ}), and then introducing the Pauli principle (\ref{sub:Approximate-treatment-of}).
The results are discussed (\ref{sub:Dependence-of-visibility}) and
compared against the experimental data (\ref{sub:Comparison-with-experiment}).
Finally, we briefly indicate (\ref{sub:Relation-to-intrinsic}) a
possible solution to the puzzling visibility oscillations observed
in the MZI without detector channel.

Our main results are: the exact formula for the time-evolution of
the visibility of the charge qubit given in equation (\ref{VisibilityMainFormula}),
and the formula for the effect of the nonequilibrium part of the noise
on the visibility of qubit (\ref{vtilde}) or interferometer (\ref{visMZ}).
Their most important general analytical consequences are derived in
section \ref{sub:Nonequilibrium-part-of}, including a detailed discussion
of the visibility oscillations.

\newcommand{\ket}[1]{\left|#1\right\rangle }

\section{Charge qubit subject to non-Gaussian shot noise}

Interferometers may be used as highly sensitive detectors, by coupling
them to a quantum system and reading out the induced phase shift.
Here we focus on a setup like the one that has been realized in \cite{2000_06_SprinzakHeiblum_ControlledDephasingOfDoubleDot},
where a double dot ({}``charge qubit'') has been subject to the
shot noise of a partitioned one-dimensional electron channel. However,
we note that the strong-coupling regime to be discussed below yet
remains to be achieved in such an experiment.

\subsection{Model}

\label{sub:Model}We consider a charge qubit with two charge states
$\hat{\sigma}_{z}=\pm1$. It is coupled to the density fluctuations
of non-interacting {}``detector'' fermions 

\begin{equation}
\hat{H}=\hat{H}_{{\rm qb}}+\hat{H}_{{\rm int}}+\hat{H}_{{\rm det}},\end{equation}
with $\hat{H}_{{\rm qb}}=\frac{\epsilon}{2}\hat{\sigma}_{z}$,

\begin{equation}
\hat{H}_{{\rm det}}=\sum_{k}\epsilon_{k}\hat{d}_{k}^{\dagger}\hat{d}_{k},\end{equation}
and

\begin{equation}
\hat{H}_{{\rm int}}=\frac{\hat{\sigma}_{z}+1}{2}\hat{V}.\label{Hint}\end{equation}
This coupling is of the diagonal form, i.e. it commutes with the qubit
Hamiltonian, thereby leading only to pure dephasing and not to energy
relaxation (the populations of the qubit levels are preserved). The
derivation of the exact expressions to be analyzed below depends crucially
on this type of coupling. The fluctuating quantum noise potential
$\hat{V}$ introduced in (\ref{Hint}) is related to the density of
detector particles in the vicinity of the qubit, see figure \ref{fig0}: 

\begin{equation}
\hat{V}=\int dx\, u(x)\hat{\rho}_{{\rm det}}(x)=\sum_{k',k}u_{k'k}\hat{d}_{k'}^{\dagger}\hat{d}_{k}\label{Vdef}\end{equation}
Here $u(x)$ is the arbitrary interaction potential (whose details
in a realistic situation will be determined by the screening properties
of the environment), $\hat{\rho}_{{\rm det}}(x)=\hat{\psi}^{\dagger}(x)\hat{\psi}(x)$
is the detector density, and $\hat{\psi}(x)=\sum_{k}\phi_{k}(x)\hat{d}_{k}$
is the expansion in terms of the single-particle eigenstates of the
detector. At this point we do not yet specify the nature of the detector,
as some of the following formulas are valid in general for any non-interacting
fermion system. However, ultimately the evaluations will be performed
for a one-dimensional channel of fermions moving chirally at constant
speed, representing our model of a {}``detector edge channel''.
This was implemented in the integer Quantum Hall Effect two-dimensional
electron gas \cite{2000_06_SprinzakHeiblum_ControlledDephasingOfDoubleDot,2006_Neder_QuantumEraser,2006_07_MZ_DephasingNonGaussianNoise_NederMarquardt}
and can presumably be realized in other one-dimensional electron systems
as well (e.g. electrons moving inside a carbon nanotube).

We are interested in describing the outcome of the following standard
type of experiment in quantum coherent dynamics: Suppose we prepare
the qubit in a superposition state of $\ket{\uparrow}$ and $\ket{\downarrow}$
at time $t=0$, and then switch on the interaction with the detector
electrons. In effect this can be realized by applying a Rabi $\pi/2$
pulse to the qubit that is initially in the state $\ket{\downarrow}$.
During the following time-evolution, the off-diagonal element $\rho_{\uparrow\downarrow}(t)$
will be affected by the coupling to the bath, experiencing decoherence.
Its original oscillatory time-evolution is multiplied by a factor,
that can be written as the overlap $D(t)=\left\langle \chi_{\downarrow}(t)|\chi_{\uparrow}(t)\right\rangle $
of the two detector states $\chi_{\downarrow}(t)$ and $\chi_{\uparrow}(t)$
that evolve under the action of $\hat{H}_{{\rm det}}$ and $\hat{H}_{{\rm det}}+\hat{V}$,
respectively. In this way, the relation between decoherence and measurement
becomes evident \cite{1990_04_SAI}:

\begin{equation}
D(t)=\left\langle e^{+i\hat{H}_{{\rm det}}t}e^{-i(\hat{H}_{{\rm det}}+\hat{V})t}\right\rangle .\label{cohfactor}\end{equation}
Note that we have set $\hbar\equiv1$. This can also be written as

\begin{equation}
D(t)=\left\langle \hat{T}\exp(-i\int_{0}^{t}dt'\,\hat{V}(t'))\right\rangle ,\label{Dtimeordered}\end{equation}
where $\hat{V}(t')$ is the fluctuating quantum noise operator in
the Heisenberg picture with respect to $\hat{H}_{{\rm det}}$, and
$\hat{T}$ is the time-ordering symbol. The magnitude of this time-dependent
{}``coherence factor'' $ $defines what we will call the {}``\emph{visibility}'' 

\begin{equation}
v=|D(t)|.\end{equation}
The visibility (with $0\leq v\leq1$) determines the suppression of
the oscillations in any observable that is sensitive to the coherence
between the two levels, e.g. $\left\langle \hat{\sigma}_{x}(t)\right\rangle ={\rm Re}\rho_{\uparrow\downarrow}(t)$.
This is depicted in figure \ref{fig0}~(d).

\subsection{Time-evolution of the visibility: General expressions}

\label{sub:Time-evolution-of-the}The average in the coherence factor
$D(t)$ displayed in equation (\ref{cohfactor}) is taken with respect
to the unperturbed state of the detector electrons, which may refer
to a nonequilibrium situation. We will assume that this initial state
can be described by independently fluctuating occupations $\hat{d}_{k}^{\dagger}\hat{d}_{k}$
of the single-particle states $k$. This assumption covers all the
cases of interest to us, namely the equilibrium noise at arbitrary
temperature, as well as shot noise produced by transmission of particles
through a partially reflecting barrier, leading to a nonequilibrium
Fermi distribution. 

The average (\ref{cohfactor}) can be evaluated in a variety of ways,
e.g. using the linked cluster expansion applied to a time-ordered
exponential. However, here we make use of a convenient formula derived
by Klich \cite{2002_Klich_Formula} in the context of full counting
statistics. Denoting as $\Gamma(A)\equiv\sum_{k',k}A_{k'k}\hat{d}_{k'}^{\dagger}\hat{d}_{k}$
the second-quantized single-particle operator built from the transition
matrix elements $A_{k'k}$, we have \cite{2002_Klich_Formula} (for
fermions)

\begin{equation}
{\rm tr}[e^{\Gamma(A)}e^{\Gamma(B)}]={\rm det}[1+e^{\hat{A}}e^{\hat{B}}],\label{KlichFormula}\end{equation}
where $\hat{A}$ is the operator acting in the \emph{single-particle}
Hilbert space. In general, this formula allows us to obtain the average
of the exponential of any single-electron operator with respect to
a many-particle density matrix that does not contain correlations.
\textbf{}Indeed, for a state with independently fluctuating occupations,
we can write the many-body density matrix in an exponential form that
is suitable for application of equation (\ref{KlichFormula}):

\begin{eqnarray}
\hat{\rho} & = & \Pi_{k}[n_{k}\hat{d}_{k}^{\dagger}\hat{d}_{k}+(1-n_{k})(1-\hat{d}_{k}^{\dagger}\hat{d}_{k})]\\
 & = & \Pi_{k}(1-n_{k})e^{\sum_{k}\hat{d}_{k}^{\dagger}\hat{d}_{k}{\rm ln}\frac{n_{k}}{1-n_{k}}},\end{eqnarray}
where $0\leq n_{k}\leq1$ is the probability of state $k$ being occupied
(formally it is necessary to consider the limits $n_{k}\rightarrow0$
and $n_{k}\rightarrow1$ if needed). Inserting this expression into
(\ref{KlichFormula}), and defining the occupation number matrix $n_{k'k}=\delta_{k'k}n_{k}$,
we are now able to evaluate averages of the form

\begin{eqnarray}
{\rm tr}[e^{i\sum_{k',k}A_{k'k}\hat{d}_{k'}^{\dagger}\hat{d}_{k}}\hat{\rho}] & = & \left(\Pi_{k}(1-n_{k})\right){\rm det}[1+e^{i\hat{A}}\frac{\hat{n}}{1-\hat{n}}]\\
 & = & {\rm det}[1+(e^{i\hat{A}}-1)\hat{n}].\end{eqnarray}
The average (\ref{cohfactor}) then can be performed by identifying
the product of time-evolution operators as a single unitary operator,
of the form given here. Thus, we find

\begin{equation}
D(t)={\rm det}[1+(\hat{S}(t)-1)\hat{n}],\label{Dgeneral}\end{equation}
where the finite-time scattering matrix (interaction picture evolution
operator) is

\begin{equation}
\hat{S}(t)=e^{i\hat{h}_{{\rm det}}t}e^{-i(\hat{h}_{{\rm det}}+\hat{u})t}.\end{equation}
Here $\hat{u}$ is the interaction from (\ref{Vdef}), and $\hat{h}_{{\rm det}}$
is the single-particle Hamiltonian of the detector electrons that
is diagonal in the $k$-basis: $\left[h_{{\rm det}}\right]_{k'k}=\epsilon_{k}\delta_{k'k}$
. In principle, equation (\ref{Dgeneral}) allows us to evaluate the
time-evolution of the coherence factor for coupling to an arbitrary
noninteracting fermion system. 

In practice, this involves calculating the time-dependent scattering
of arbitrary incoming $k$-states from the coupling potential $u(x)$,
i.e. determining the action of the scattering matrix. Note that in
the case of fully occupied states ($n_{k}\equiv1$ for all $k$),
the operator $\hat{n}$ becomes the identity and the determinant reduces
to the product of scattering phase factors that can be obtained by
diagonalizing the scattering matrix. More generally, the contributions
to $D(t)$ from states deep inside the Fermi sea always only amount
to a phase factor, which will drop out when considering the visibility
$v=|D(t)|$. 

In the remainder of this paper, we will focus on the specific, and
experimentally relevant, case of a one-dimensional channel of fermions
moving at constant speed $v_{{\rm det}}$ (i.e. using a linearized
dispersion relation). We will employ plane wave states inside a normalization
volume $L$ and first assume a finite bandwidth $k\in[-k_{c},+k_{c}]$.
At the end of the calculation, we will send $L$ and $k_{c}$ to infinity
(see below).

The equation of motion for a detector single-particle wave function
$\psi(x,t)$ in the presence of the potential $u(x)$ is

\begin{equation}
i(\partial_{t}+v_{{\rm det}}\partial_{x})\psi(x,t)=u(x)\psi(x,t),\end{equation}
which is solved by 

\begin{equation}
\psi(x,t)=\exp[-i\int_{0}^{t}dt'\, u(x-v_{{\rm det}}t')]\psi(x-v_{{\rm det}}t,0).\end{equation}
This corresponds to the action of $\exp(-i(\hat{h}_{{\rm det}}+\hat{u})t)$
on the initial wave function. Applying $\exp(i\hat{h}_{{\rm det}}t)$
afterwards, we end up with the same expression, but with $\psi(x,0)$
on the right-hand-side (rhs). In other words, the action of the scattering
matrix is to multiply the wave function by a position-dependent phase
factor:

\begin{equation}
\left[\hat{S}(t)\psi\right](x)=e^{-iw(x)}\psi(x),\label{Sequation}\end{equation}
where the phase function $w(x)$ is related to the interaction potential,
as seen above:

\begin{equation}
w(x)=\int_{0}^{t}dt'\, u(x-v_{{\rm det}}t').\label{wx}\end{equation}
The phase function is depicted in figure \ref{fig0}~(c). Two remarks
regarding the finite band-cutoff $k_{c}$ are in order at this point:
As argued above, states deep inside the Fermi sea only contribute
a phase factor to $D(t)$. This is the reason we obtain a converging
result for the visibility $v=|D(t)|$ when taking the limit $k_{c}\rightarrow\infty$
in the end, whereas $D(t)$ itself acquires a phase that grows linearly
with $k_{c}$. Moreover, strictly speaking the relation (\ref{Sequation})
only holds for states $\psi(x)$ that are not composed of $k$-states
at the boundaries of the interval $k\in[-k_{c},+k_{c}]$, since otherwise
the multiplication by $e^{-iw(x)}$ will yield contributions that
are cut off as they fall outside the range of allowed wavenumbers.
Nevertheless, for the purpose of calculating the visibility, this
discrepancy between the operators $\hat{S}(t)$ and $e^{iw(x)}$ will
not matter, as those states only contribute phases to $D(t)$ anyway.
Thus, we are indeed allowed to write the visibility as

\begin{equation}
v(t)=\left|{\rm det}[1+(\hat{S}(t)-1)\hat{n}]\right|=\left|{\rm det}[1+(e^{-i\hat{w}(t)}-1)\hat{n}]\right|.\label{VisibilityMainFormula}\end{equation}
This is the central formula that will be the basis for all our discussions
below.

We briefly discuss some general properties of the phase function $w(x)$
and its Fourier transform. The matrix elements of $\hat{w}$ are given
by the Fourier transform $w_{k'k}=\frac{1}{L}\tilde{w}(q=k'-k)$ of
$w(x)$. Thus, they are connected to those of the interaction potential
$u(x)$ via

\begin{equation}
\tilde{w}(q)=\int dx\, e^{-iqx}w(x)=\frac{e^{iqv_{{\rm det}}t}-1}{iqv_{{\rm det}}}\tilde{u}(q),\label{wq}\end{equation}
where $\tilde{u}(q)=\int dxe^{-iqx}u(x)$. 

At times $v_{{\rm det}}t\gg\sigma$, the phase function $w(x)$ has
the generic form of a box with corners rounded on the scale $\sigma$
of the interaction potential, see figure \ref{fig0}~(c). The phase
fluctuations are then due to the fluctuations of the number of electrons
inside the interval of length $v_{\det}t$. The most important parameter
in this regard is the height of $w(x)$ inside the interval. This
defines the \emph{dimensionless coupling strength} $g$, given by

\begin{equation}
g\equiv\frac{\tilde{w}(q=0)}{v_{{\rm det}}t}=\frac{1}{v_{{\rm det}}}\int_{-\infty}^{+\infty}dy\, u(y).\label{gdef}\end{equation}
The coupling strength determines the contribution of a single electron
to the phase (in a regime where we are allowed to treat that single
electron simply as a delta peak in the density). We will see that
all the results can be expressed in terms of the dimensionless quantities
$g,\, eVt$, $V\sigma/v_{{\rm det}}$, and the occupation probability
$\mathcal{T}$ of states inside the voltage window (as well as the
temperature, $T\sigma/v_{{\rm det}}$, for finite temperature situations).

\subsection{Relation to full counting statistics}

\label{sub:Relation-to-full}In the context of full counting statistics
(FCS) \cite{1996_06_LevitovLeeLesovik_FullCountingStatistics}, one
is interested in obtaining the entire probability distribution of
a fluctuating number $N$ of particles, e.g. the number of electrons
transmitted through a certain wire cross section during a given time
interval, or the number of particles contained within a certain volume.
Usually, it is most convenient to deal with the generating function

\begin{equation}
\chi(\lambda)=\sum_{N}P_{N}e^{i\lambda N}.\end{equation}
The decoherence function $D(t)$, and thus the visibility $v=|D(t)|$,
are directly related to a suitably defined generating function. In
the limit $\sigma\rightarrow0$, the phase function $w(x)$ becomes
a box of height $g$ on the interval $x\in[0,v_{{\rm det}}t]$. Then
$\int dx\, w(x)\hat{\rho}(x)$ is $g\hat{N}$, where $\hat{N}$ is
the number of electrons within the box. Thus we find for the visibility

\begin{equation}
v=|\chi(g)|,\end{equation}
in terms of the generating function $\chi$ for the probability distribution
of particles $N$. For a finite range $\sigma$ of the interaction
potential, we are dealing with a fluctuating quantity that no longer
just takes discrete values. 

We emphasize, however, that our main focus is different from the typical
applications of FCS, where one is usually interested in the long-time
limit and consequently discusses the remaining small deviations from
purely Gaussian statistics. The long-time behaviour of decoherence
by a detecting quantum point contact has been discussed in \cite{2005_09_AverinSukhorukov_FCS_QPCdetectorPRL},
where formulas similar to (\ref{VisibilityMainFormula}) appeared.
In contrast, we are interested in the visibility oscillations as a
most remarkable feature of the behaviour at short to intermediate
times. In other words, the kinds of setups discussed here in principle
offer an experimental way of accessing such short-time features of
FCS, which are otherwise not detectable.

\subsection{Gaussian approximation}

\label{sub:Gaussian-approximation}Before going on to discuss the
visibility arising from the exact expression (\ref{VisibilityMainFormula}),
we derive the Gaussian approximation to the visibility. We will first
do so in a general way and later point out that the same result could
be obtained starting from equation (\ref{VisibilityMainFormula}).
If $\hat{V}$ were a linear superposition of harmonic oscillator coordinates,
and these oscillators were in thermal equilibrium, then its noise
would be Gaussian (i.e. it would correspond to a Gaussian random process
in the classical limit). This kind of quantum noise, arising from
a harmonic oscillator bath, is the one studied most of the time in
the field of quantum dissipative systems (e.g. in the context of the
Caldeira-Leggett model or the spin-boson model). In that case, the
following expression would be exact. In contrast, our present model
in general displays non-Gaussian noise, being due to the density fluctuations
of a system of discrete charges. Thus, the following formula constitutes
what we will call the {}``Gaussian approximation'', against which
we will compare the results of our model:

\begin{equation}
D_{{\rm Gauss}}(t)=\exp\left[-i\left\langle \hat{V}\right\rangle t-\frac{1}{2}\int_{0}^{t}dt_{1}\int_{0}^{t}dt_{2}\left\langle \hat{T}\delta\hat{V}(t_{1})\delta\hat{V}(t_{2})\right\rangle \right].\end{equation}
Here $\delta\hat{V}\equiv\hat{V}-\left\langle \hat{V}\right\rangle $.
If we are only interested in the decay of the visibility, we obtain

\begin{equation}
v_{{\rm Gauss}}(t)=|D_{{\rm Gauss}}(t)|=\exp\left[-\frac{1}{2}\int_{0}^{t}dt_{1}\int_{0}^{t}dt_{2}\frac{1}{2}\left\langle \left\{ \delta\hat{V}(t_{1}),\delta\hat{V}(t_{2})\right\} \right\rangle \right],\end{equation}
i.e. the decay only depends on the symmetrized quantum correlator.
Introducing the quantum noise spectrum

\begin{equation}
\left\langle \delta\hat{V}\delta\hat{V}\right\rangle _{\omega}=\int dt\, e^{i\omega t}\left\langle \delta\hat{V}(t)\delta\hat{V}(0)\right\rangle ,\end{equation}
we find the well-known expression

\begin{equation}
\ln v_{{\rm Gauss}}(t)=-\int\frac{d\omega}{2\pi}\left\langle \delta\hat{V}\delta\hat{V}\right\rangle _{\omega}\frac{2\sin^{2}(\frac{\omega t}{2})}{\omega^{2}}.\end{equation}
This result is valid for an arbitrary noise correlator. Inserting
the relation between $\hat{V}$ and the density fluctuations (\ref{Vdef}),
we have

\begin{equation}
\left\langle \delta\hat{V}\delta\hat{V}\right\rangle _{\omega}=2\pi\sum_{k',k}\left|u_{k'k}\right|^{2}n_{k}(1-n_{k'})\delta(\omega-(\epsilon_{k'}-\epsilon_{k}))\end{equation}
for the spectrum. In the following, we specialize to the case of one-dimensional
fermions moving at constant speed ($\epsilon_{k}=v_{{\rm det}}k$).
Then we find:

\begin{equation}
{\rm ln}v_{{\rm Gauss}}(t)=-\frac{1}{2}\sum_{k',k}|w_{k'k}(t)|^{2}n_{k}(1-n_{k'}),\label{Gaussian}\end{equation}
where the matrix $\hat{w}$ corresponds to the potential $u(x-v_{{\rm det}}t')$
integrated over the interaction time, see equations (\ref{wx}) and
(\ref{wq}). We could have arrived at this formula equally well by
using (\ref{VisibilityMainFormula}) to write

\begin{equation}
v=\exp[{\rm Re}\,{\rm tr}\ln(1+(e^{-i\hat{w}}-1)\hat{n})],\end{equation}
and expanding the exponent to second order in $\hat{w}$. Equation
(\ref{Gaussian}) will be used for comparison against the full results
obtained from (\ref{VisibilityMainFormula}) numerically below.

\subsection{Results for the visibility according to the Gaussian approximation}

\label{sub:Results-for-the}We now discuss the results of the Gaussian
approximation for certain special cases. For definiteness, here and
in the following, we will assume an interaction potential $u(x)$
of width $\sigma$, which we will take to be of Gaussian form wherever
the precise shape is needed:

\begin{equation}
u(x)=\frac{u_{0}}{\sqrt{\pi}\sigma}e^{-(x/\sigma)^{2}},\,\,\tilde{u}(q)=u_{0}e^{-(q\sigma/2)^{2}}.\label{ushape}\end{equation}
Other smoothly decaying functions do not yield results that deviate
appreciably in any qualitatively important way. The coupling strength
(\ref{gdef}) then becomes

\begin{equation}
g=\frac{u_{0}}{v_{{\rm det}}}.\end{equation}
At zero temperature, in equilibrium, the evolution of the visibility
is determined by the well-known physics of the orthogonality catastrophe,
which underlies many important phenomena such as the X-ray edge singularity
or the Kondo effect \cite{2000_Mahan}: After coupling the two-level
system to the fermionic bath, the two states $\ket{\chi_{\uparrow}}$
and $\ket{\chi_{\downarrow}}$ evolve such that their overlap decays
as a power-law. The long-time limit of a vanishing overlap is produced
by the fact that the ground states of a fermion system with and without
an arbitrarily weak scattering potential are orthogonal. The exponent
can be obtained from the coupling strength $g$. We find, from (\ref{Gaussian})
and (\ref{wq}),

\begin{equation}
v_{{\rm Gauss}}^{T=V=0}(t)={\rm const}\cdot\left(\frac{v_{{\rm det}}t}{\sigma}\right)^{-\left(\frac{g}{2\pi}\right)^{2}}\label{powerlaw}\end{equation}
in the long-time limit. Only the prefactor depends on the precise
shape of the interaction $u(x)$. We note that the result diverges
for $\sigma\rightarrow0$. The reason is that a finite $1/\sigma$
is needed as an effective momentum cutoff up to which fluctuations
of the density in the Fermi sea are taken into account. In any physical
realization the fluctuations will be finite, since then the density
of electrons is finite and there is a physical cutoff besides $1/\sigma$.

After applying a finite bias voltage, the occupation inside the voltage
window is determined by the transmission probability $\mathcal{T}$
of a barrier (quantum point contact) through which the stream of electrons
has been sent: $n_{k}=\mathcal{T}$ for $v_{{\rm det}}k\in[0,eV]$.
Then, equation (\ref{Gaussian}) yields two contributions, one of
which is the equilibrium contribution we have just calculated. As
a result, the visibility \emph{factorizes} into the zero voltage contribution
and the extra suppression resulting from the second moment of the
shot noise:

\begin{equation}
\frac{v_{{\rm Gauss}}^{T=0,V\neq0}(t)}{v_{{\rm Gauss}}^{T=0,V=0}(t)}=\exp\left[-\mathcal{T}(1-\mathcal{T})\left(\frac{g}{2\pi}\right)^{2}F(eVt)\right]\label{Factorization}\end{equation}
The function $F(eVt)$ is given by

\begin{equation}
F(eVt)=\left\{ \begin{array}{l}
\frac{(eVt)^{2}}{2}\,\,(eVt\ll1\,\,{\rm and}\,\, eV\sigma/v_{{\rm det}}\ll1)\\
\pi eVt\,\,(eVt\gg1,\sigma=0)\end{array}\right.\label{Ffunc}\end{equation}
Here the low-voltage (short-time) quadratic rise is independent of
the shape of the interaction potential $u$: Only low frequency (long
wavelength) fluctuations of the density are important, and thus only
the coupling constant $g$ enters, being an integral over $u(x)$,
see (\ref{gdef}). At large voltages there is, in general, an extra
constant prefactor in front of $F$ that depends on $\sigma$ and
the shape of $u$. However, in contrast to the equilibrium part of
the visibility (\ref{powerlaw}), the limit $\sigma\rightarrow0$
is finite, and we have evaluated this limit in the second line of
(\ref{Ffunc}).

Finally, it is interesting to note that for the present model the
fermionic density can be expressed as a sum over normal mode oscillators
(plasmons) after bosonization. Thus, in equilibrium, the Gaussian
approximation is actually exact. However, once the system is driven
out of equilibrium by a finite bias voltage and displays shot noise,
the many-body state is a highly correlated non-Gaussian state, when
expressed in terms of the plasmons, even though it looks simple with
respect to the fermion basis, where the occupations of different $k$-states
fluctuate independently.

\subsection{Exact numerical results for the visibility and discussion}

\label{sub:Exact-numerical-results}In the following, we plot and
discuss the results of a direct numerical evaluation of the determinant
(\ref{VisibilityMainFormula}) that yields the exact time-evolution
of the visibility of a charge qubit subject to shot noise. We focus
on the zero temperature case, although the formula also allows us
to treat thermal fluctuations which lead to an additional suppression
of visibility. The most important parameters are the dimensionless
coupling constant $g$ (\ref{gdef}), the transmission probability
$\mathcal{T}$, and the voltage $V$ applied to the detector channel.
We will also note whenever the results depend on the width $\sigma$
or shape of the interaction potential $u(x)$. 

\begin{figure}
\includegraphics[%
  width=0.49\columnwidth]{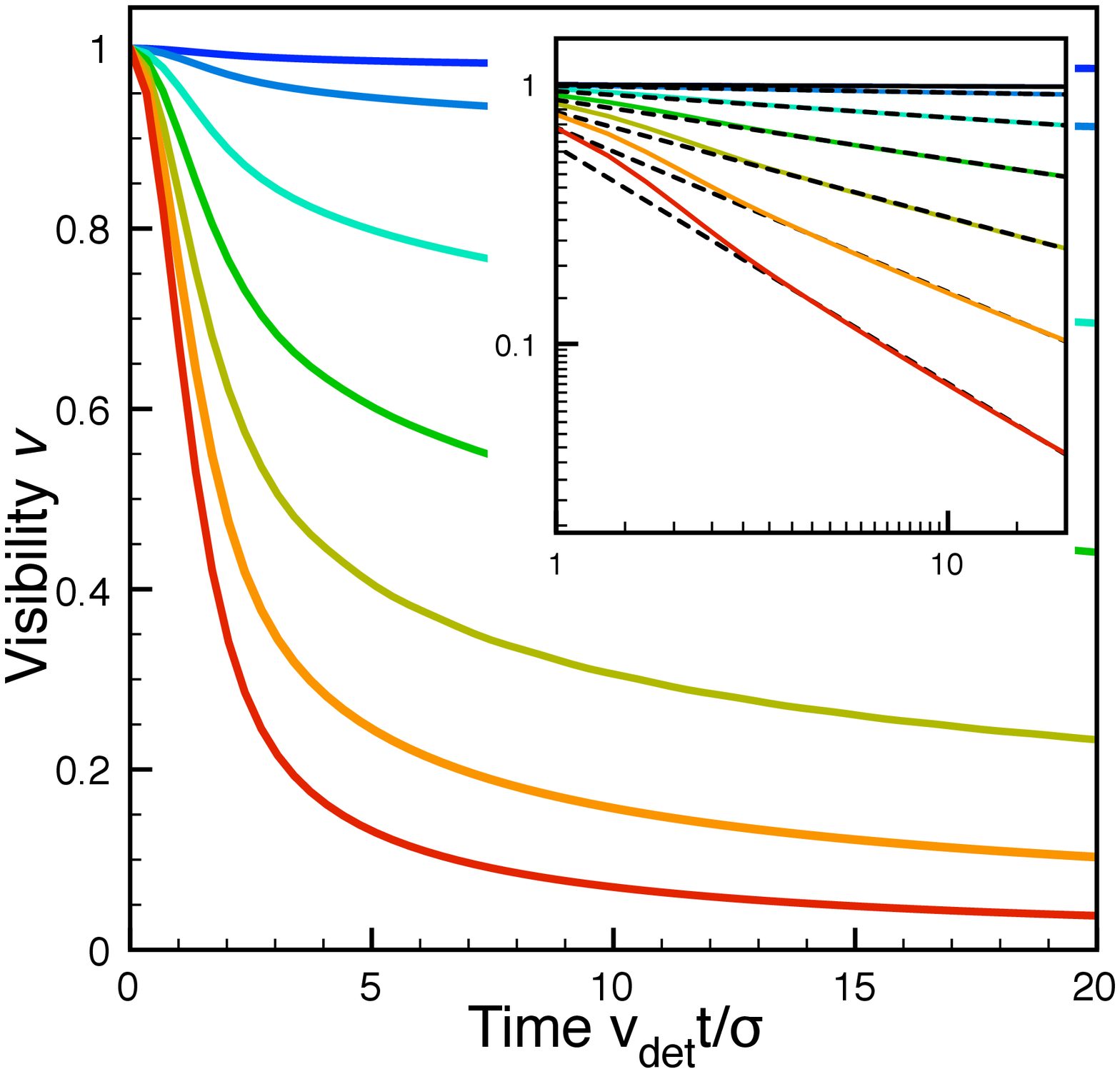}~\includegraphics[%
  width=0.49\columnwidth]{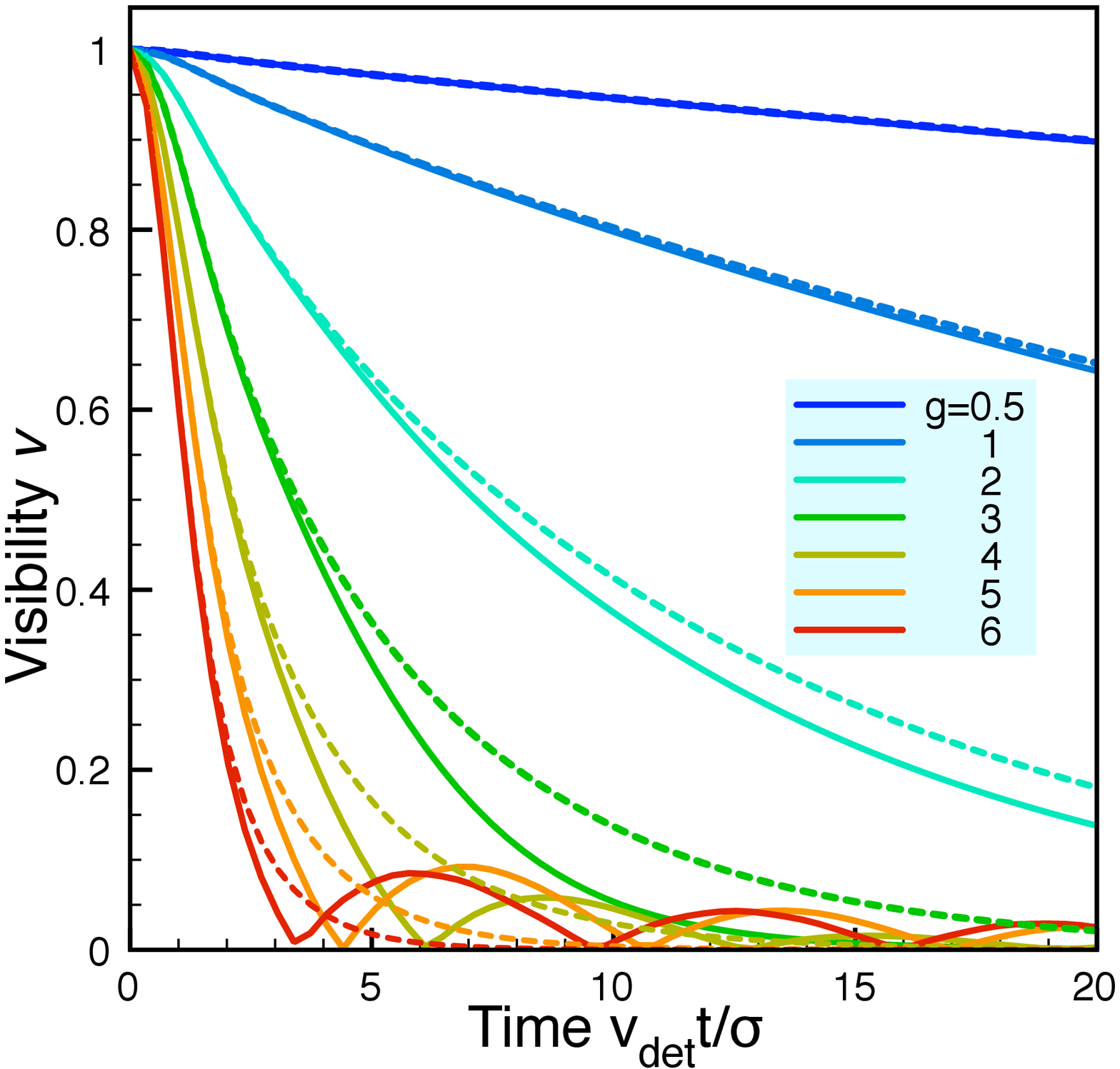}

\caption{\label{fig1}Time-evolution of the visibility (coherence) $v=|D(t)|$
of a qubit coupled to quantum noise from a non-interacting 1D electron
channel, at zero temperature, after switching on the interaction at
$t=0$. The curves have been obtained by direct numerical evaluation
of (\ref{VisibilityMainFormula}). \emph{Left}: Decoherence by equilibrium
noise, for increasing coupling strength $g$ (top to bottom curve),
displaying the power-law decay (\ref{powerlaw}) expected from the
physics of the orthogonality catastrophe (inset: log-log plot, with
dashed lines indicating the expected exponents $(g/2\pi)^{2}$). \emph{Right}:
Decoherence by shot-noise (at a finite voltage $eV\sigma/v_{{\rm det}}=1$).
Beyond $g=\pi$, the visibility displays a periodic pattern, with
zeroes and coherence revivals, as a result of the non-Gaussian nature
of the noise. Dashed lines indicate the Gaussian approximation. The
transmission probability of the barrier generating the shot noise
equals $\mathcal{T}=1/2$. }
\end{figure}

In figure \ref{fig1}, we have displayed the time-evolution of the
visibility $v=|D(t)|$ as a function of $v_{{\rm det}}t/\sigma$,
for different couplings. In equilibrium, the curves derived from the
full expression (\ref{VisibilityMainFormula}) coincide exactly with
those obtained from the Gaussian theory (\ref{Gaussian}), as expected.
The long-time behaviour is given by the power-law decay (\ref{powerlaw})
arising from the orthogonality catastrophe. However, at finite voltages,
with extra dephasing due to shot noise, the Gaussian approximation
fails: In general, it tends to overestimate the visibility at longer
times and larger couplings (dashed lines in figure \ref{fig1},~right).
The most prominent non-Gaussian feature sets in after the coupling
$g$ crosses a threshold that is equal to $g=\pi$, as will be explained
below: For larger $g$, the visibility displays oscillations, vanishing
at certain times (for a barrier with $\mathcal{T}=1/2$) and showing
{}``coherence revivals'' in-between these zeroes. The zeroes coincide
with phase jumps of $\pi$ in $D(t)$ (see figure \ref{fig:TransmissionVaried}).
We will discuss the locations of these zeroes in more detail below.

Such a behaviour of the visibility can \emph{only} be explained by
invoking non-Gaussian noise. In every Gaussian theory, we can employ
$\left\langle e^{i\varphi}\right\rangle =e^{-\left\langle \varphi^{2}\right\rangle /2}\geq0$,
which directly excludes the behaviour found here (regardless of noise
spectrum and coupling strength), even though it is still compatible
with a non-monotonous evolution of the visibility. The simplest available
model of dephasing by non-Gaussian noise will compared with the present
results in section \ref{sub:Comparison-with-dephasing}.

\begin{figure}
\begin{center}\includegraphics[%
  width=0.8\columnwidth]{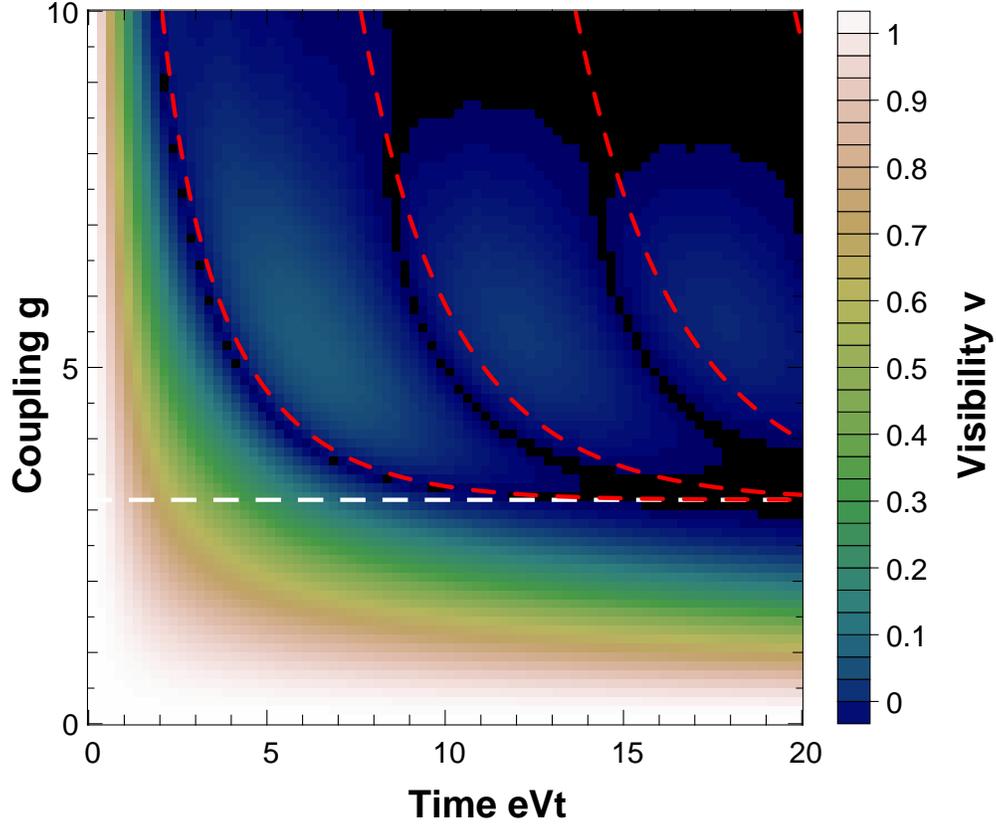}\end{center}

\caption{\label{fig2}Evolution of the visibility (density plot) as a function
of coupling constant $g$ (vertical) and time $eVt$ (horizontal).
Visibility oscillations start beyond $g=\pi$. Note the unequal spacing
between zeroes: The first zero occurs at a time $eVt_{1}=2\pi^{2}/g$
for large $g\gg1$, see (\ref{firstzero}). The spacing of subsequent
zeroes is given approximately by $\delta(eVt)=2\pi$ for the regime
of couplings considered here. Further parameters: $\mathcal{T}=1/2$
and $eV\sigma/v_{{\rm det}}=1$. The red dashed lines indicate the
expected location of visibility zeroes, according to the approximation
$v'(t)$ for the nonequilibrium part, (\ref{vtilde}), in the limit
$\sigma\rightarrow0$. }
\end{figure}

In order to obtain insight into the general structure of the solution,
we first of all note that the qualitative features (in particular
the zeroes of the visibility) depend only weakly on the width $\sigma$
or shape of the interaction potential. In fact, these features are
due to the non-equilibrium part of the noise, and the Gaussian approximation
suggests that the limit $\sigma\rightarrow0$ is well-defined for
that part. This is the reason why, in the following, we will plot
the time-evolution as a function of $eVt$ (instead of $v_{{\rm det}}t/\sigma$),
which is the relevant variable. 

\begin{figure}
\begin{center}\includegraphics[%
  width=0.48\columnwidth]{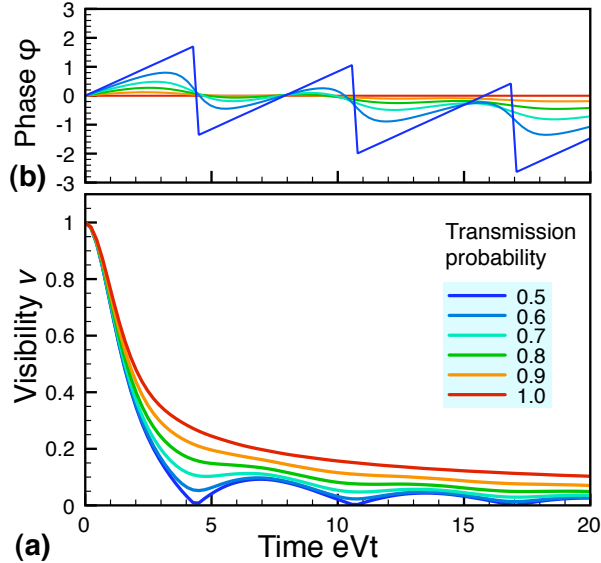}\end{center}

\caption{\label{fig:TransmissionVaried}Effect of the transmission probability
and phase evolution. (\textbf{a}) Visibility $v=|D(t)|$ as a function
of time for different transmission probabilities $\mathcal{T}$ (where
$g=5$ and $eV\sigma/v_{{\rm det}}=1$). (\textbf{b}) Corresponding
evolution of the phase, i.e. the argument of the complex coherence
factor $D_{r}(t)$, where the subscript indicates that the phase evolution
for $\mathcal{T}=1$ has been subtracted as a reference. }
\end{figure}

In figure \ref{fig2}, we display the time-evolution versus the coupling
$g$. The threshold at $g=\pi$ is clearly noted. Furthermore, the
first zero occurs at a time $t_{1}=2\pi^{2}/(eVg)$ which shrinks
with increasing coupling (provided $g\gg1$, see discussion in the
next section and equation (\ref{firstzero})). In contrast, subsequent
zeroes have a periodic spacing that appears to be roughly independent
of $g$, given approximately by $\delta(eVt)=2\pi$ for the small
values of $g$ plotted here. 

Finally, in figure \ref{fig:TransmissionVaried}, the effects of the
transmission probability $\mathcal{T}$ on the evolution of the visibility
and the full coherence factor $D(t)$ have been plotted, indicating
the phase jumps obtained at $\mathcal{T}=1/2$ whenever $v=|D(t)|$
vanishes. 

All of these features will now be analyzed further by restricting
the discussion to the effects of the nonequilibrium part of the noise.

\section{Nonequilibrium part of the noise}

\label{sec:Nonequilibrium-part-of}

\subsection{General properties}

\label{sub:Nonequilibrium-part-of}Within the Gaussian approximation,
we noted that the visibility at finite voltages factorized into one
factor describing the decay due to equilibrium noise and another part
describing the effect of nonequilibrium shot noise (\ref{Factorization}).
More precisely, the nonequilibrium part of the visibility can be calculated
from (\ref{Gaussian}) by simply restricting the matrix elements of
$w_{k'k}$ to transitions within the voltage window: $k,k'\in[0,eV/v_{{\rm det}}]$.
This has the physical interpretation that only these transitions contibute
to the excess noise in the spectrum $\left\langle \delta\hat{V}\delta\hat{V}\right\rangle _{\omega}$
of the fluctuating potential. In addition, since the equilibrium noise
comes out exact in the Gaussian theory, we can state that all the
non-Gaussian features are due to the nonequilibrium part.

Based on these observations, we now introduce a heuristic approximation
to the full \emph{non-Gaussian} theory, which works surprisingly well.
We will factorize 

\begin{equation}
v^{T=0,V\neq0}(t)\approx v^{T=V=0}(t)\cdot v'(t),\label{vapproxFactor}\end{equation}
where $v'$ is the visibility obtained from the full expression (\ref{VisibilityMainFormula})
after restricting the matrix elements of $\hat{w}$ in the fashion
described above. We will denote the restricted matrix as $\hat{w}'$.
Note that the restricted matrix depends on the voltage, in contrast
to $\hat{w}$ itself. 

\begin{figure}
\includegraphics[%
  width=0.32\columnwidth]{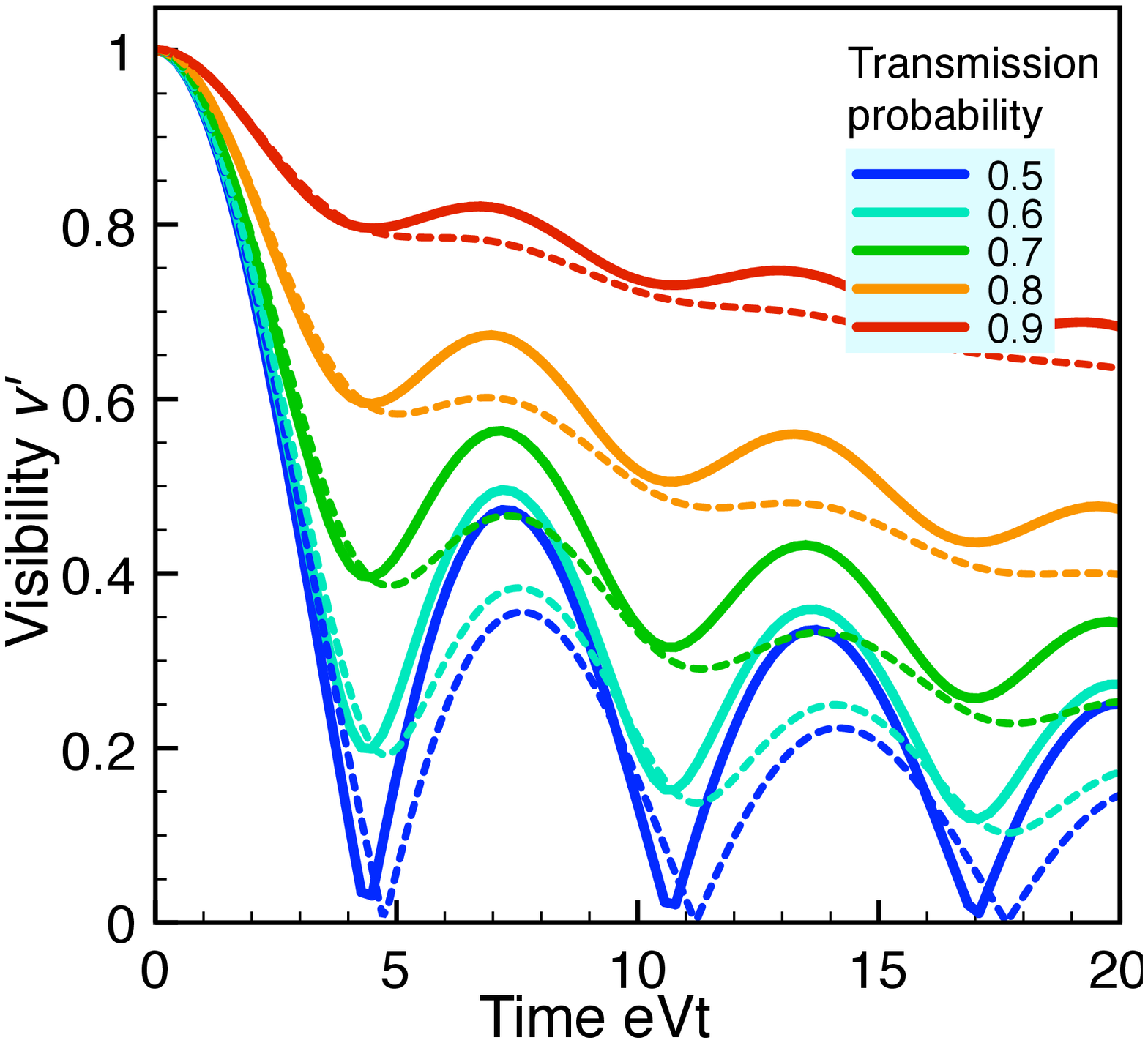}~\includegraphics[%
  width=0.32\columnwidth]{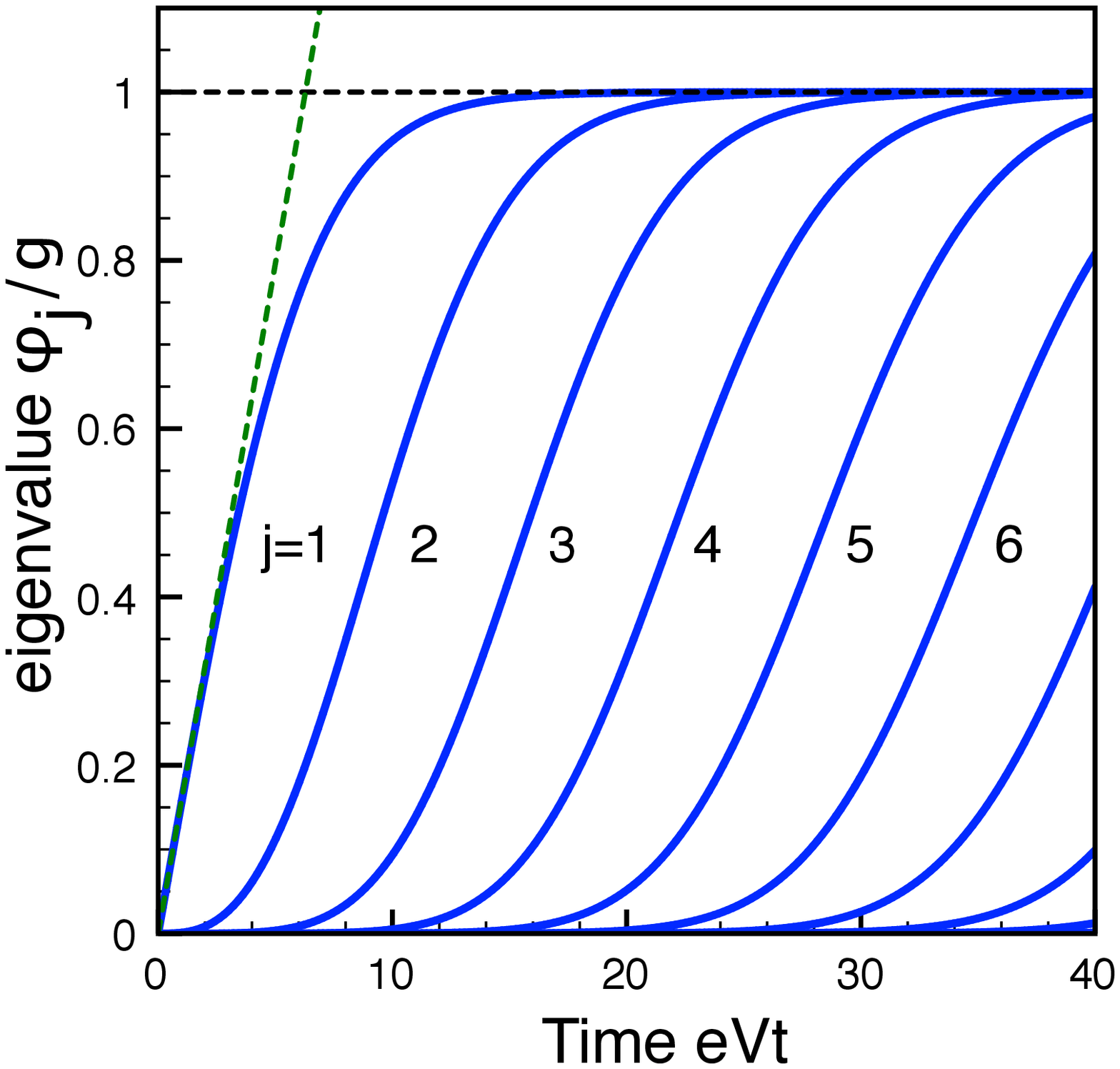}~\includegraphics[%
  width=0.32\columnwidth]{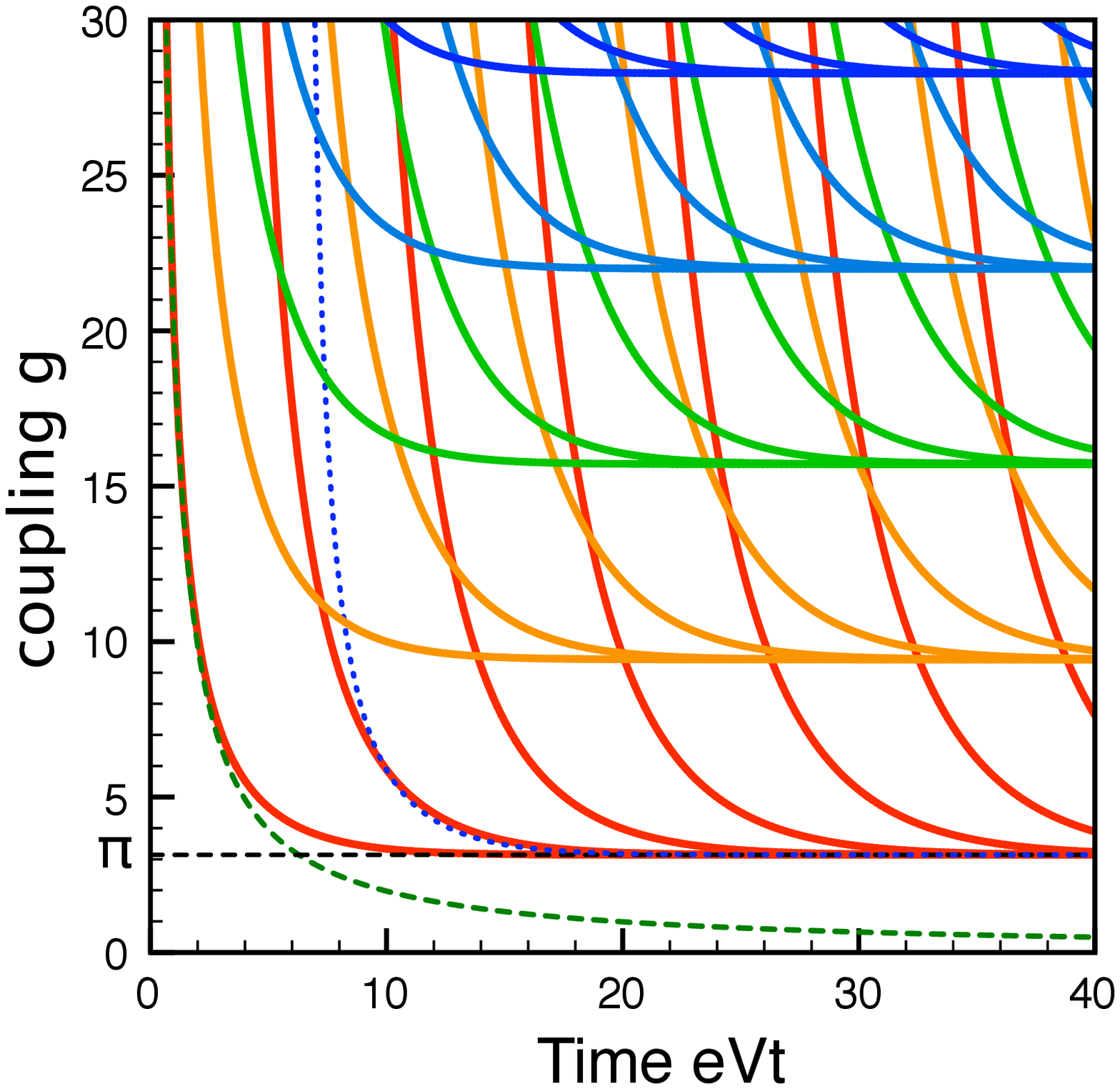}

\caption{\emph{\label{vtildefig}Left}: Contribution $v'$ of the nonequilibrium
part of the noise (i.e. the shot noise) to the suppression of the
visibility. The full lines have been obtained from the exact result
(\ref{VisibilityMainFormula}) by dividing by the result at zero voltage,
$v^{V\neq0}(t)/v^{V=0}(t)$. The dashed lines represent the approximation
of restricting matrix elements of $\hat{w}$ to the voltage window,
see (\ref{vtilde}). We have set $eV\sigma/v_{{\rm det}}=1$ in this
panel. \emph{Middle}: Universal curves for the eigenvalues $\varphi_{j}$
of the restricted matrix $\hat{w}'$ entering the visibility $v'(t)$
in equation (\ref{vtilde}), plotted as a function of $eVt$ in the
limit $\sigma\rightarrow0$. \emph{Right}: Locations of the zeroes
in the visibility $v'(t)$ as a function of coupling $g$ (compare
figures \ref{fig2} and \ref{fig3}). These curves can be obtained
from those on the left by taking $(2n+1)\pi/(\varphi_{j}/g)$, with
$n=0,1,2,3,4,\ldots$ (from bottom to top). The blue dotted line corresponds
to the first curve displaced by $2\pi$, indicating the periodicity
observed for small couplings $g$. In both panels, the green dashed
line shows the short-time behaviour $\varphi_{1}=geVt/2\pi$.}
\end{figure}

Since the occupation probability is constant within the voltage window,
$n_{k}=\mathcal{T}$, the matrices $\hat{n}$ and $\hat{w}'$ now
commute. This allows a considerable simplification, yielding a visibility
that can be written in terms of the eigenvalues $\varphi_{j}$ of
the matrix $\hat{w}'$:

\begin{equation}
v'(t)=\Pi_{j}|\mathcal{R}+\mathcal{T}e^{-i\varphi_{j}}|.\label{vtilde}\end{equation}
Thus, the dependence on the transmission probability has been separated
from the dependence on interaction potential, time, and voltage, contained
within $\varphi_{j}$. The results obtained from the exact formula
are compared against this approximation in figure \ref{vtildefig}~(a).
We observe that all the important qualitative features are retained
in the approximation. Furthermore, the locations of the zeroes come
out quite well, while the amplitude of the oscillations is underestimated.

We will now list some general properties of the matrix $\hat{w}'$
that determines the visibility according to (\ref{vtilde}):\\
(i) The sum of eigenvalues is 

\begin{equation}
\sum_{j}\varphi_{j}={\rm tr}\hat{w}'=\frac{g}{2\pi}eVt.\label{sumrule}\end{equation}
(ii) For a non-negative (non-positive) phase function $w(x)$, the
matrix $\hat{w}'$ is positive (negative) semidefinite: We can map
any wavefunction $\ket{\psi}$ to another state $\ket{\psi'}$ by
setting $\psi'_{k}=\psi_{k}$ only inside the voltage window, and
$\psi'_{k}=0$ otherwise. Then

\begin{equation}
\left\langle \psi\left|\hat{w}'\right|\psi\right\rangle =\left\langle \psi'\left|\hat{w}\right|\psi'\right\rangle =\int dx\,|\psi'(x)|^{2}w(x)\geq0,\label{nonnegative}\end{equation}
for a non-negative function $w(x)$, and analogously for a non-positive
function $w(x)$.\\
(iii) Following the same argument, we can prove that the largest eigenvalue
of $\hat{w}'$ is bounded by the maximum of $w(x)$, if ${\rm max}w(x)\ge0$:

\begin{equation}
\left\langle \psi\left|\hat{w}'\right|\psi\right\rangle =\int dx\,|\psi'(x)|^{2}w(x)\leq\left\langle \psi'|\psi'\right\rangle {\rm max}w(x)\leq{\rm max}w(x).\end{equation}
Analogously the smallest eigenvalue is bounded from below by the minimum
(if ${\rm max}w(x)\leq0$).

At small voltages (short times) (where $eVt\ll1$ and $eV\sigma/v_{{\rm det}}\ll1$),
the matrix elements are constant inside the voltage window, $w'_{k'k}=\frac{1}{L}\tilde{w}(q=0)$,
yielding only one nonvanishing eigenvalue, given by $ $(\ref{sumrule}):

\begin{equation}
\varphi_{1}=\frac{g}{2\pi}eVt\,.\label{onlyeigenvalue}\end{equation}
As a consequence, at sufficiently large $g\gg1$, the first zero in
the visibility $v'(t)$ will occur when $\varphi_{1}=\pi$, implying

\begin{equation}
t_{1}=\frac{2\pi^{2}}{geV}.\label{firstzero}\end{equation}
Assuming now that $w(x)$ is non-negative (as is the case in our example,
if $g>0$), we can immediately deduce the following general consequences
from properties (i) to (iii): All of them taken together imply that
the rise of the first eigenvalue must saturate below ${\rm max}w(x)$
(which approaches $g$ for times $v_{{\rm det}}t\gg\sigma$). Thus,
other eigenvalues must start to grow, in order to obey the sum-rule.
If (and only if) the coupling constant is large enough, this may lead
to an infinite series of zeroes in the visibility (see below). Therefore,
we are dealing with a true strong coupling effect.

We have not found an analytical way of obtaining $\varphi_{j}$ at
arbitrary parameters. However, all relevant features follow from the
foregoing discussion and may be illustrated by numerical evaluation
of the eigenvalues.

\begin{figure}
\begin{center}\includegraphics[%
  width=0.8\columnwidth]{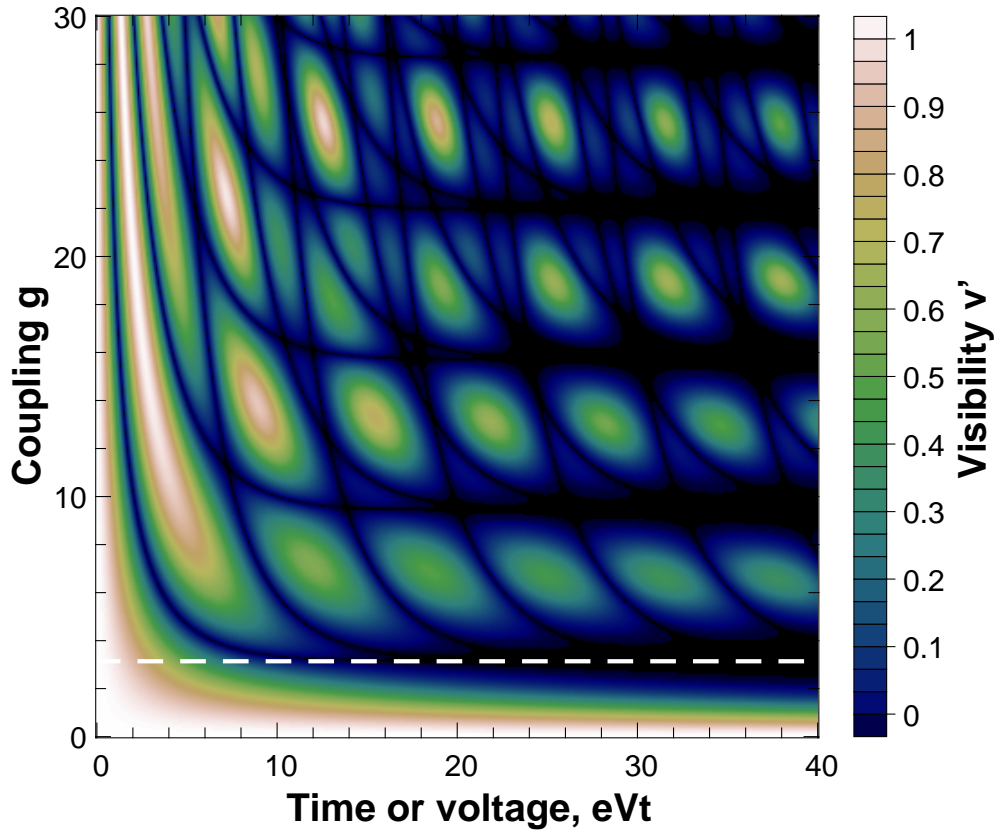}\end{center}

\caption{\label{fig3}Nonequilibrium part of the visibility, $v'$, as a function
of coupling strength $g$ and time (or voltage) $eVt$. The lines
of vanishing $v'$ are those plotted in figure \ref{vtildefig}. The
structure will be distorted for $\sigma>0$, and the full visibility
$v$ will be further suppressed at higher couplings by the dephasing
due to equilibrium noise, see (\ref{vapproxFactor}). }
\end{figure}
We note that the limit $\sigma\rightarrow0$ is well-defined, and
we will assume this limit in the following, in which results become
independent of the shape of the interaction potential. This limit
represents a good approximation as soon as the time is sufficiently
large: $v_{{\rm det}}t\gg\sigma$. In that limit, the eigenvalues
have the following functional dependence:

\begin{equation}
\varphi_{j}=g\cdot\varphi_{j}^{(g=1)}(eVt).\end{equation}
Thus the complete behaviour at all coupling strengths can be inferred
by numerically evaluating the eigenvalues once as a function of $eVt$.
This has been done in figure \ref{fig3}. 

At $\mathcal{T}=1/2$, the visibility $v'(t)$ will vanish whenever
one of these eigenvalues is equal to $(2n+1)\pi$, where $n=0,1,2,\ldots$.
Thus, the locations of the zeroes can be obtained from the equation
$(2n+1)\pi=\varphi_{j}$, or equivalently $g=(2n+1)\pi/\varphi_{j}^{(g=1)}$.
The latter equation has the advantage that the rhs is independent
of $g$. It has been used in the right panel of figure \ref{fig3}.
These curves have also been inserted into figure \ref{fig2}, for
comparison against the results from the full theory. In particular,
the first zero is reproduced very accurately, while there are some
quantitative deviations at subsequent zeroes.

The full pattern of the visibility, as a function of interaction time
and coupling strength, can become very complex due to the large number
of lines of vanishing visibility $v'$. This is depicted in figure
\ref{fig3}.

The analysis in the next section indicates (and the numerical results
displayed in figure \ref{fig3} confirm) that the spacing between
the subsequent zeroes in the visibility is no longer determined by
$g$, but rather given by $2\pi/eV$ (with deviations at higher $g$).
As we will explain in the next section, this corresponds to one additional
detector electron passing by the qubit during the interaction time.

\subsection{Wave packet picture}

\label{sub:Wave-packet-picture}Following Martin and Landauer \cite{1992_01_MartinLandauer_WavepacketApproach},
we introduce a new basis of states inside the voltage window, whose
width in k-space is set by $\Delta k\equiv eV/v_{{\rm det}}$:

\begin{equation}
\ket{\psi_{n}}=\left(\frac{\Delta k}{2\pi/L}\right)^{-1/2}\sum_{k\in[0,\Delta k]}e^{-ikn\Delta x}\ket{k}.\end{equation}
In real space, these states represent a train of wavepackets, spaced
apart by $\Delta x=2\pi/\Delta k=v_{{\rm det}}\Delta t$, corresponding
to time bins of duration $\Delta t=h/eV=2\pi/eV$. Taking the limit
$L\rightarrow\infty$, we have

\begin{equation}
\psi_{n}(x)=\sqrt{\frac{\Delta k}{2\pi}}e^{i\frac{\Delta k}{2}(x-n\Delta x)}{\rm sinc}(\frac{\Delta k}{2}(x-n\Delta x)),\end{equation}
where ${\rm sinc}(y)=\sin(y)/y$. These packets move at constant velocity,
$\psi_{n}(x,t)=\psi_{n}(x-v_{{\rm det}}t)$. Their advantage is that
they are localized in space and therefore well suited for calculating
matrix elements of the phase function $w(x)$ (or its restricted counterpart
$\hat{w}'$). From

\begin{equation}
\left\langle \psi_{n'}\left|\hat{w}\right|\psi_{n}\right\rangle =\left(\frac{\Delta k}{2\pi/L}\right)^{-1}\sum_{k,k'\in[0,\Delta k]}e^{i(k'n'-kn)\Delta x}w_{k'k}\end{equation}
we find

\begin{equation}
\left\langle \psi_{n'}\left|\hat{w}\right|\psi_{n}\right\rangle =\frac{\Delta k}{2\pi}\int_{-1}^{+1}d\tilde{q}\, e^{i\pi(n+n')\tilde{q}}G_{n'-n}(|\tilde{q}|)\tilde{w}(\tilde{q}\Delta k),\end{equation}
where

\begin{equation}
G_{n'-n}(|\tilde{q}|)=\left\{ \begin{array}{c}
\frac{\sin(\pi|\tilde{q}|(n'-n))}{\pi(n-n')}\,\,{\rm for}\, n'\neq n\\
1-|\tilde{q}|\,\,{\rm for}\, n'=n\end{array}\right..\end{equation}
These formulas enable a very efficient numerical evaluation. In general,
for large $|n|$ and $|n'|$ the corresponding wave packets lie outside
the range of $w(x)$ and therefore the corresponding matrix elements
are small and can be neglected. In the limit of small voltages, the
matrix $\hat{w}'$ is already diagonal in this basis (compare (\ref{onlyeigenvalue})):

\begin{equation}
\left\langle \psi_{n'}\left|\hat{w}\right|\psi_{n}\right\rangle \approx\left[\frac{geVt}{2\pi}\right]\delta_{n'0}\delta_{n0}.\end{equation}
Moreover, the wave packet basis permits a very intuitive interpretation
of the results in the Gaussian approximation: First, let us assume
that the number of contributing wave packets is large, $N\equiv eVt/2\pi\gg1$.
The wave packets are orthonormalized, and the phase function $w(x)$
is smooth on the scale $\Delta x=v_{{\rm det}}\Delta t=v2\pi/eV$
of these packets (for sufficiently large voltages). As a result, we
find that the matrix $\hat{w}$ is diagonal in this basis, up to terms
of order $1/N$:

\begin{equation}
\left\langle \psi_{n'}\left|\hat{w}\right|\psi_{n}\right\rangle \approx\delta_{n'n}w(x=n\Delta x)+O\left(\frac{1}{N}\right).\end{equation}
Therefore, in any sum over the eigenvalues $\varphi_{j}$, these can
be approximated by the values of $w$ taken at the centers of the
wave packets. Assuming further that the coupling $g\ll1$ is weak,
each of the $\varphi_{j}$ is small. This allows us to expand the
visibility reduction due to nonequilibrium noise:

\begin{equation}
v'(t)=\Pi_{j}|\mathcal{R}+\mathcal{T}e^{-i\varphi_{j}}|\approx1-\frac{1}{2}\mathcal{RT}\sum_{j}\varphi_{j}^{2}+\ldots\end{equation}
That is the expected result, which has the form of {}``phase diffusion'',
with a contribution from the variance of the phase shift exerted by
each detector electron. In the limit of $\sigma\ll eVt$, we get $\varphi_{j}=g$
for approximately $N$ wave packets, and zero otherwise. Then we reproduce
equation (\ref{Ffunc}) in the long-time limit:\begin{equation}
\sum_{j}\varphi_{j}^{2}=Ng^{2}=\frac{t}{\Delta t}g^{2}=\frac{eVt}{2\pi}g^{2}.\label{constantgphi2}\end{equation}
More generally, for any shape of $w(x)$ we can replace 

\begin{equation}
\sum_{j}\varphi_{j}^{2}\approx\sum_{n}\left\langle \psi_{n}\left|\hat{w}\right|\psi_{n}\right\rangle ^{2}\approx\frac{1}{\Delta x}\int dx\, w^{2}(x).\end{equation}
Defining the effective width of $w(x)$ as

\begin{equation}
l_{{\rm eff}}\equiv\frac{\left[\int dx\, w(x)\right]^{2}}{\int dx\, w^{2}(x)},\label{leff}\end{equation}
we can set the total number of wave packets to be $N=l_{{\rm eff}}/\Delta x$.
Defining the average phase shift induced by a single detector electron
as $\bar{\varphi}=\int dx\, w(x)/(N\Delta x)=(geVt/2\pi)/N$, we can
write

\begin{equation}
\sum_{j}\varphi_{j}^{2}\approx N\bar{\varphi}^{2}.\end{equation}
Note that in the case of a constant $w(x)$, we have $l_{{\rm eff}}=v_{{\rm det}}t$,
and therefore $N=eVt$ and $\bar{\varphi}=g$, so we are back to (\ref{constantgphi2}).

\section{Comparison with dephasing by classical telegraph noise}

\label{sub:Comparison-with-dephasing}In this section, we compare
and contrast the results we have obtained in the rest of this paper
against the simplest possible model displaying non-Gaussian features
in dephasing: pure dephasing of a qubit by classical random telegraph
noise. There, visibility oscillations are observed once the coupling
strength becomes large as compared to the switching rate of the two-state
fluctuator producing the telegraph noise. In the limit of a vanishing
switching rate, the visibility in that model is the average of two
oscillatory phase factors evolving at different frequencies, corresponding
to the two different energy shifts imparted by the two-state fluctuator:

\begin{equation}
v^{{\rm tel.noise}}(t)=|(1-p)+pe^{-i\delta\omega t}|.\end{equation}
If the occupation probability of the two fluctuator states is $p=1/2$,
this leads to visibility oscillations $v^{{\rm tel.noise}}(t)=\left|\cos(\delta\omega t)\right|$,
roughly similar to those found in our full quantum theory of dephasing
by shot noise. The decaying envelope of these oscillations is then
produced by a finite switching probability. 

It is instructive to set up a rough correspondence between that simple
model and the one considered here, and see how far it takes us (and
where it fails): According to the well-known semiclassical picture
of binomial shot noise \cite{2000_BlanterBuettiker_ReviewSN}, during
the time-interval $\Delta t=h/eV=2\pi/eV$ a single detector electron
arrives with a probability $\mathcal{T}$. It imparts a phase shift
$g$ within our model. Thus, the fluctuator probability $p$ would
equal $\mathcal{T}$, the mean time between telegraph noise switching
events would be taken as $\Delta t$, and the frequency difference
$\delta\omega$ would have to be set equal to $g/\Delta t$. This
analogy partly suggests the right qualitative behaviour, namely a
threshold in $g$ that is independent of voltage (independent of $\Delta t$).
This threshold turns out to be $g=1$ in the telegraph noise model,
and for larger $g$ the visibility oscillates with a period $4\pi\Delta t/\sqrt{g^{2}-1}$
(if $p=\mathcal{T}=1/2$). Although this correctly suggests that the
first zero occurs at a position $t\propto1/(eVg)$, it predicts all
subsequent zeroes to occur at the same period, which is not compatible
with the actual behaviour (see figure \ref{fig2}). These discrepancies
are not too surprising, since the two models certainly differ even
qualitatively in the following sense: In random telegraph noise, the
switching occurs in a Markoff process, i.e. without memory. In contrast,
in the semiclassical model of binomial shot noise the electrons arrive
in a stream of \emph{regularly} spaced time-bins of size $\Delta t=h/eV$.
We have not found any reasonable way of incorporating this fact into
a simplified semiclassical model, since it is unclear how to treat
'fractional time-bins' within such a model.

\section{Electronic Mach-Zehnder interferometer coupled to a detector edge
channel}

\label{sec:Electronic-Mach-Zehnder-interferometer}In this section
we will show how to explain the surprising experimental results that
have been obtained recently in a strongly coupled {}``which-path
detector system'' involving a Mach-Zehnder interferometer coupled
to a {}``detector'' edge channel. We will present a nonperturbative
treatment that captures all the essential features due to the non-Gaussian
nature of the detector shot noise. Our approximate solution for this
model is directly related to the exact solution of the simpler charge
qubit system discussed above.

A simplified scheme of the experimental setup is presented in figure
\ref{fig0} (see \cite{2006_Neder_QuantumEraser,2006_07_MZ_DephasingNonGaussianNoise_NederMarquardt}
for a detailed explanation). Both the MZI and the detector were realized
utilizing chiral one-dimensional edge-channels in the integer Quantum
Hall effect regime. The MZI phase was controlled by a modulation gate
via the Aharonov-Bohm (AB) effect. The additional edge channel was
partitioned by a quantum point contact, before traveling in close
proximity to the upper path of the MZI, serving as a \char`\"{}which
path\char`\"{} phase-sensitive detector \cite{2000_06_SprinzakHeiblum_ControlledDephasingOfDoubleDot}.
For a finite bias applied to the detector channel, the Coulomb interaction
between both channels caused orbital entanglement between the interfering
electron and the detecting electrons, thereby decreasing the contrast
of the AB oscillations. This contrast, quantified in terms of the
\emph{visibility} $v=(I_{{\rm max}}-I_{{\rm min}})/(I_{{\rm max}}+I_{{\rm min}})$,
was measured as a function of the DC bias $V$ applied at the detector
channel, and of the partitioning probability $\mathcal{T}$ of the
detector channel. 

As noted already in the introduction, two new and peculiar effects,
which will be explained here, were observed in these experiments:

\begin{itemize}
\item \emph{Unexpected dependence on partitioning}: The visibility as a
function of $\mathcal{T}$ changes from the expected smooth parabolic
suppression $\propto\mathcal{T}(1-\mathcal{T})$ at low detector voltages
to a sharp {}``V-shape'' behaviour at some larger voltages, with
almost zero visibility at $\mathcal{T}=1/2$ (see Fig. 3 in \cite{2006_07_MZ_DephasingNonGaussianNoise_NederMarquardt}).
\item \emph{Visibility oscillations}: For some values of the detector QPC
gate voltage (yielding $\mathcal{T}\approx1/2$), the visibility drops
to zero at intermediate voltages, then reappears again as $V$ increases,
in order to vanish at even larger voltages (see Fig. 4 in \cite{2006_07_MZ_DephasingNonGaussianNoise_NederMarquardt}).
For some other gate voltages it decreases monotonically (see Fig.
2 in \cite{2006_Neder_QuantumEraser}).
\end{itemize}
In \cite{2006_07_MZ_DephasingNonGaussianNoise_NederMarquardt} we
showed that a simplified model involving a single detector electron
can provide a qualitative explanation for the experimental results
listed above. However, it has clear shortcomings, both quantitative
and in terms of the physical interpretation. The natural reason for
these shortcomings is that detection in the experiment is due to a
varying number of electrons, not just a single one. Then two questions
arise: (i) How many electrons dephase the MZI as the detector voltage
increases, and (ii) how much does each electron contribute to dephasing.
These questions will be answered by the following model.

\subsection{Solution of the single-particle problem}

\label{sub:ModelMZ}The main simplifying assumption in our approach
will be that it is possible to treat each given electron in the Mach-Zehnder
interferometer on its own, as a single particle interacting with the
fluctuations of the density in the detector channel. Making this assumption
is far from being a trivial step, as it effectively neglects Pauli
blocking, and we will have to comment on it in the next section. For
now, however, let us define the following model as our starting point:

\begin{equation}
\hat{H}=v_{{\rm MZ}}\hat{p}+\int dx'\, u(x'-\hat{x})\hat{\rho}_{{\rm det}}(x')+\hat{H}_{{\rm det}}.\label{HamLin}\end{equation}
Here $\hat{x}$ and $\hat{p}=-i\partial_{x}$ are the position and
the momentum operator, respectively, of the single interfering electron
under consideration (traveling in the upper, interacting path of the
interferometer). We have linearized the dispersion relation, keeping
in mind that the interferometer's visibility will be determined by
the electrons near the MZ Fermi energy, traveling at a speed $v_{{\rm MZ}}$.
The Fermi energy itself has been subtracted as an irrelevant energy
offset, and likewise the momentum is measured with respect to the
Fermi momentum. The Aharonov-Bohm phase between the interfering paths
would have to be added by hand.

We thus realize that the situation is analogous to the model treated
above, involving pure dephasing of a charge qubit. The two states
of the qubit correspond to the two paths which the interfering electron
can take. The following analysis explicitly demonstrates this equivalence
and arrives at an expression for the visibility which is the analogue
of equation (\ref{Dtimeordered}). The only difference will be the
replacement of $v_{{\rm det}}$ by the relative velocity $v_{{\rm det}}-v_{{\rm MZ}}$,
which can be understood by going into the frame of reference of the
MZ electron. 

Let us now consider the full wave function $\ket{\Psi_{{\rm total}}(t)}=e^{+i\hat{H}_{{\rm det}}t}e^{-i\hat{H}t}\ket{\Psi_{{\rm total}}(0)}$
of MZI and detector, expressed in the interaction picture with respect
to $\hat{H}_{{\rm det}}$. One can always decompose the full wave
function in the form $\ket{\Psi_{{\rm total}}(t)}=\int dx\,\ket{x}\otimes\ket{\psi(x,t)}$.
$ $Here, we focus on the projection $\ket{\psi(x,t)}\equiv\left\langle x\left|\Psi_{{\rm total}}(t)\right.\right\rangle $
onto the MZI single-particle position basis. This is a state in the
detector Hilbert space, with $\left\langle \left.\psi(x,t)\right|\psi(x,t)\right\rangle $
giving the probability of the MZI electron to be found at position
$x$. It obeys the Schr\"odinger equation

\begin{equation}
i\frac{\partial}{\partial t}\ket{\psi(x,t)}=\left[-iv_{{\rm MZ}}\frac{\partial}{\partial x}+\hat{V}(x,t)\right]\ket{\psi(x,t)},\label{eqmotionpsi}\end{equation}
where the fluctuating potential $\hat{V}(x,t)\equiv\int dx'\, u(x'-x)\hat{\rho}_{{\rm det}}(x',t)$
is in the interaction picture with respect to $\hat{H}_{{\rm det}}$.
The \emph{exact} solution of equation (\ref{eqmotionpsi}) that follows
from the Hamiltonian (\ref{HamLin}) reads:

\begin{equation}
\ket{\psi(x,t)}=\hat{T}\exp\left[-i\int_{0}^{t}dt'\,\hat{V}(x-v_{{\rm MZ}}t',t-t')\right]\ket{\psi(x-v_{{\rm MZ}}t,0)}.\end{equation}
Thus, at a given space-time point $(x,t)$, the {}``quantum phase''
in the exponent is an integral over the values of the fluctuating
potential at all points on the {}``line of influence'' $(x',t')$
with $x-x'=v_{{\rm MZ}}(t-t')$. If the potential were classical,
the exponential would represent a simple phase factor. Here, however,
the interfering electron is not only acted upon by the fluctuations
but also changes the state of the detector. The state $\ket{\psi(x,t)}$
contains all the information about the entanglement between the MZI
electron and the detector electrons. 

We will now determine the visibility resulting from the interaction
between interferometer and detector channel. At the first beam splitter,
the electron's wave packet is decomposed into two parts, one of them
traveling along the lower ($l$) arm of the interferometer, the other
one traveling along the upper ($u$) arm. These are described by states
$\ket{\psi_{l}(x,t)}$ and $\ket{\psi_{u}(x,t)}$, respectively, which
obey the Schr\"odinger equation given above, albeit in general with
a different noise potential for each of them. The visibility is determined
by the overlap between those two states, taken at the position $x=v_{{\rm MZ}}t$
of the second beam splitter (where $t$ is the time-of-flight through
the interferometer):

\begin{equation}
v=\left|\left\langle \psi_{l}(x,t)|\psi_{u}(x,t)\right\rangle \right|.\end{equation}
Taking into account that $\ket{\psi_{l,u}(0,0)}=\ket{\Psi_{{\rm det}}}$
is the detector's initial unperturbed state, and realizing that the
interaction takes place only in the upper arm, we find that the visibility
is determined by the probability amplitude for the electron to exit
the MZI without having changed the state of the detector \cite{1990_04_SAI}:

\begin{equation}
v=\left|\left\langle \Psi_{{\rm det}}\left|\hat{T}\exp\left[-i\int_{0}^{t}dt'\,\hat{V}(x-v_{{\rm MZ}}t',t-t')\right]\right|\Psi_{{\rm det}}\right\rangle \right|.\label{visMZ1}\end{equation}
The initial detector state $\ket{\Psi_{{\rm det}}}$ itself is produced
by partitioning a stream of electrons. 

The last step consists in representing equation (\ref{visMZ1}) as
an expectation value of a unitary operator:

\begin{equation}
v=\left|\left\langle \Psi_{{\rm det}}\left|e^{-i\hat{\Phi}}\right|\Psi_{{\rm det}}\right\rangle \right|,\label{visMZexpPhi}\end{equation}
where $\hat{\Phi}$ is defined as the operator in the exponent of
(\ref{visMZ1}). We have been allowed to drop the time ordering symbol
because the density fluctuations in the one-dimensional detector channel
are described by free bosons: $\left[\hat{\rho}(x,t),\hat{\rho}(x',t')\right]$
is a purely imaginary c-number. The time-ordered exponential is by
definition a product of many small unitary evolutions sorted by time.
Hence, using repeatedly the Baker-Hausdorff formula $e^{\hat{A}}e^{\hat{B}}=e^{\hat{A}+\hat{B}}e^{[\hat{A},\hat{B}]/2}$,
which holds since $[\hat{A},\hat{B}]$ commutes with $\hat{A}$ and
$\hat{B}$ in this case, we can collect the operators at different
times into the same exponent. The remaining c-number exponent only
contributes a phase, so it does not lead to a reduction in the visibility
and we can disregard it.

The phase operator $\hat{\Phi}$ in (\ref{visMZexpPhi}) is therefore
a weighted integral over the density operator:

\begin{equation}
\hat{\Phi}=\int_{0}^{t}dt'\hat{V}(x-v_{{\rm MZ}}t',t-t')=\int dx'\, w(x')\hat{\rho}_{{\rm det}}(x')dx'=\sum_{k,k'}w_{k'k}\hat{d}_{k'}^{\dagger}\hat{d}_{k}.\label{Phiexpr}\end{equation}
The phase function $w(x)$ is the one that has been introduced before,
in equation (\ref{wx}), with the exception that the detector velocity
has to be replaced by the \emph{relative} velocity: $v_{{\rm det}}\mapsto v_{{\rm det}}-v_{{\rm MZ}}$.
It can be viewed as a convolution of the interaction potential $u(x)$
with the {}``window of influence'' of length $l=|(v_{{\rm det}}-v_{{\rm MZ}})t|$
defined by the traversal time $t$ and the velocities.

\subsection{Approximate treatment of Pauli blocking}

\label{sub:Approximate-treatment-of}The loss of visibility is due
to the trace a particle leaves in the detector \cite{1990_04_SAI}.
If the detector (or, in general, the environment) is in its ground
state initially, this means that the detector has to be left in an
excited state afterwards. Energy conservation implies that the energy
has to be supplied by the particle itself. This is no problem if the
particle starts out in an excited state. An example is provided by
a qubit in a superposition of ground and excited state, which can
decay to its ground state by spontaneous emission of radiation into
a zero-temperature environment. 

However, in electronic interference experiments such as the one considered
here, we are interested in the loss of visibility with regard to the
interference pattern observed in the \emph{linear} conductance. At
zero temperature, this implies we are dealing with electrons right
at the Fermi surface which have no phase space available for decay
into lower-energy states, due to Pauli blocking. Only an environment
that is itself in a nonequilibrium state (e.g. the voltage-biased
detector channel) can then lead to dephasing. This very basic physical
picture has been confirmed by many different calculations. While it
is, in principle, conceivable that subtle non-perturbative effects
might eventually lead to a break-down of this picture, we are not
aware of any unambiguous and uncontroversial theoretical derivation
of a suppression of linear conductance visibility at zero temperature,
for an interferometer coupled to an equilibrium quantum bath.

The main difficulty in dealing with an electronic interferometer coupled
to a quantum bath thus lies in the necessity of treating the full
many-body problem. Any model that considers only a single interfering
particle subject to the environment will miss the effects of Pauli
blocking, and thereby permit unphysical, artificial dephasing by spontaneous
emission events that would be absent in a full treatment. In \cite{2004_10_Marquardt_MZQB_PRL,2006_04_MZQB_Long},
it was shown how to properly incorporate these effects into an equations-of-motion
approach similar to the one described above (with the fermion field
$\hat{\psi}(x,t)$ taking the role of the single-particle state $\ket{\psi(x,t)}$).
The main idea was that the state of the detector, and therefore the
noise potential $\hat{V}$, will itself be influenced by the density
in the interferometer, leading to {}``backaction terms'' (known
from the quantum Langevin equation for quantum dissipative systems)
that ultimately ensure Pauli blocking. However, in order to be able
to solve the equations of motion of the environment, it was crucial
to assume Gaussian quantum noise, and even then the solution for the
visibility was carried out only to lowest order in the coupling. Thus,
this approach is not feasible for the present problem, where we want
to keep non-Gaussian effects in a fully nonperturbative way. Nevertheless,
the underlying intuitive physical picture remains valid: If both the
interferometer and the detector are near their ground states, the
interfering electron will get {}``dressed'' by distorting the detector
electron density in its vicinity, but this perturbation is undone
when it leaves the interaction region. Therefore no trace is left
and there is no contribution to the dephasing rate.

We therefore resort to an approximate treatment (applicable to the
zero-temperature situation), suggested by the general physical picture
described above. We will continue to use the single-particle picture
for the interferometer, but keep only the nonequilibrium part of the
noise, thus eliminating the possibility of artificial dephasing for
the case when the detector is not biased. In fact, within a lowest-order
perturbative calculation, this scheme gives exactly the right answers:
Firstly, dephasing by the quantum equilibrium noise of the detector
channel is completely eliminated by Pauli blocking, as follows from
the analysis of \cite{2004_10_Marquardt_MZQB_PRL,2006_04_MZQB_Long}
(at $T=0$, for the linear conductance). Secondly, the remaining nonequilibrium
part of the noise spectrum, corresponding to the shot noise, is symmetric
in frequency, and thus equivalent to purely classical noise whose
effects are not diminished by Pauli blocking (see discussions in \cite{2006_04_MZQB_Long,2006_04_DecoherenceReview}). 

The analysis in section \ref{sub:Nonequilibrium-part-of} has demonstrated
that all the non-Gaussian features are due to the nonequilibrium part
which we retain. Therefore, we expect that the present approximation
should be able to reproduce the novel features observed in the experiment,
which is confirmed by comparison with the experimental data. \textbf{}We
emphasize once more that it is crucial to supplement the single-particle
picture by taking care of the Pauli principle afterwards.

Thus, we shall restrict the matrix elements in equation (\ref{Phiexpr})
to the voltage window, replacing $w_{k'k}$ by the restricted $w'_{k'k}$,
according to the notation introduced in section \ref{sub:Nonequilibrium-part-of}.
All that remains to be done to calculate the visibility is diagonalizing
the operator $\hat{\Phi}$, which is achieved by switching to the
basis of eigenstates of $w'$,

\begin{equation}
\hat{\Phi}=\sum_{j}\varphi_{j}\hat{c}_{j}^{\dagger}\hat{c}_{j},\end{equation}
where $\varphi_{j}$ are the eigenvalues and $\hat{c}_{j}$ is the
annihilation operator for eigenstate $j$ of $w'$. The occupation
operators $\hat{c}_{j}^{\dagger}\hat{c}_{j}$ fluctuate independently,
and all states $j$ have the same occupation probability $\mathcal{T}$,
just like the states in the original basis. This is a consequence
of the occupation matrix being proportional to the identity matrix,
as pointed out near equation (\ref{vtilde}). Therefore, equation
(\ref{Phiexpr}) reduces to

\begin{equation}
v'=\left|\left\langle e^{-i\hat{\Phi}}\right\rangle \right|=\left|\left\langle \Psi_{{\rm det}}\left|\Pi_{j}e^{-i\varphi_{j}\hat{c}_{j}^{\dagger}\hat{c}_{j}}\right|\Psi_{{\rm det}}\right\rangle \right|=\Pi_{j}\left|\mathcal{R}+\mathcal{T}e^{-i\varphi_{j}}\right|.\label{visMZ}\end{equation}
This formula is our main result for the visibility of the MZI, valid
at zero temperature. It gives a closed expression for the reduction
of the interference contrast in the AB oscillations of the MZI, as
a function of detector bias and partitioning probability $\mathcal{T}$.
It has been calculated nonperturbatively within the approximation
discussed above, i.e. employing a single-particle picture for the
interfering electron and simultaneously retaining only the nonequilibrium
part of the detector noise. At any given detector voltage $V$, there
exists a basis of states in the detector, which, when occupied, contribute
to the MZI phase by different amounts $\varphi_{j}$. These occupations
fluctuate due to the partitioning at the detector beam splitter. The
visibility then is the product of all those influences.

\subsection{Dependence of visibility on detector voltage and detector partitioning}

\label{sub:Dependence-of-visibility}

\begin{figure}
\begin{center}\includegraphics[%
  width=0.7\columnwidth]{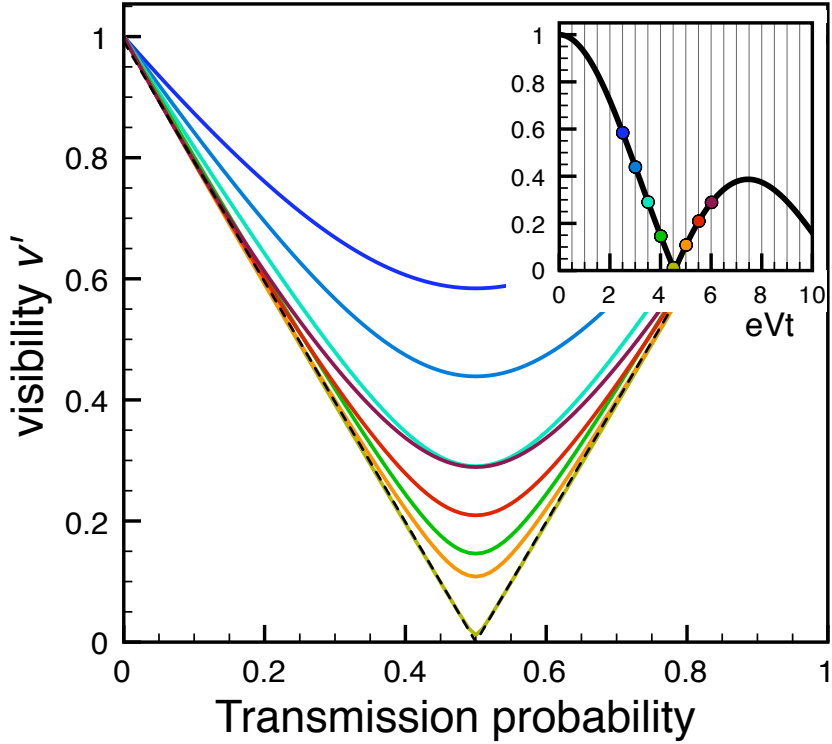}\end{center}

\caption{\label{fig4_VisibilityVsTransmission}Dependence of the visibility
$v'$ on the partitioning probability $\mathcal{T}$ of the detector
current, for different voltages, using equation (\ref{visMZ}). The
{}``V-shape'' is clearly observed. \emph{Inset}: Visibility at $\mathcal{T}=1/2$
as a function of $eVt$. The locations corresponding to the curves
in the main plot are indicated. Other parameters: $g=5$ and temperature
$T=0$. \textbf{}}
\end{figure}

The visibility for the Mach-Zehnder interferometer subject to the
shot noise in the detector channel may thus be calculated in the same
manner as the visibility for the charge qubit treated above, if the
restriction to the nonequilibrium part of the noise is taken into
account. The main difference is that in the MZI the interaction time
$t$ is dictated by the setup. However, in the limit $\sigma\rightarrow0$,
the visibility $v'$ only depends on the product $eVt$. Thus the
plots above (figures \ref{vtildefig}~(b,c) and \ref{fig3}) also
depict the dependence of $v'$ on the voltage at fixed time $t$. 

At small bias voltages, only one eigenvalue is nonzero and grows linearly
with detector voltage, according to equation (\ref{onlyeigenvalue}):
$\varphi_{1}=geVt/2\pi\equiv\gamma V$. Thus, the visibility is

\begin{equation}
v'=|\mathcal{R}+\mathcal{T}e^{-i\gamma V}|,\label{OneDetectingElectron}\end{equation}
where the proportionality constant $\gamma$ may be measured from
the voltage-dependent phase shift obtained for the non-partitioned
case, $\mathcal{T}=1$. Equation (\ref{OneDetectingElectron}) represents
the influence of {}``exactly one detecting electron''. 

One can obtain (\ref{OneDetectingElectron}) as an ansatz, by postulating
that exactly one detector electron interacts with the interfering
electron \cite{2006_07_MZ_DephasingNonGaussianNoise_NederMarquardt}.
Here we obtained it naturally as a limiting case of our full expression.
It has to be emphasized that this result is highly counterintuitive:
Naively, one would assume that each detector electron induces a constant
phase shift that is set by the coupling strength and does not depend
on the detector voltage. The voltage $V$ should only control the
frequency at which detector electrons are injected. However, to the
extent that we identify each eigenvalue $\varphi_{j}$ with one detector
electron, we have to conclude that this naive picture is wrong. Formally,
only a single, very extended detector wavepacket of size $\propto V^{-1}$
interacts with the quantum system, i.e. the charge qubit or the interfering
electron (see section \ref{sub:Wave-packet-picture}). As the interaction
range is fixed and limited, this means that the phase shift (basically
the expectation value of $w(x)$ in terms of this wave packet) then
shrinks with $V$. The linear dependence of the phase shift on voltage
thus may be rationalized by taking into account energy conservation:
At lower detector voltages $V$ the phase space for scattering of
detector electrons gets restricted severely, and thus the \emph{effective}
interaction strength is diminished. Likewise, the spatial resolution
of this which-path detector becomes very poor, as is apparent from
the large extent of the wave packet: Detection at a high spatial resolution
would prepare a localized state that contains a lot of energy, more
than is available in the detector-interferometer system.

Regarding the dependence on the transmission probability of the detector
channel (figure \ref{fig4_VisibilityVsTransmission}), we note that
there are strong deviations from the smooth dependence $\exp[-{\rm C}\mathcal{T}(1-\mathcal{T})]$
expected for any Gaussian noise model (where $C$ would depend on
$V,t,g,\ldots$ but not on $\mathcal{T}$). These deviations are particularly
strong near the voltages for which the visibility becomes zero at
$\mathcal{T}=1/2$. Indeed, if only one eigenvalue contributes and
is equal to $\pi$, equation (\ref{OneDetectingElectron}) yields
a {}``V-shape'' of the visibility, $v'=|1-2\mathcal{T}|$, as indicated
by the dashed line in figure \ref{fig4_VisibilityVsTransmission}.

\subsection{Comparison with experiment}

\label{sub:Comparison-with-experiment}In this section we briefly
discuss the results obtained by fitting the present model to the experimental
data. This follows our discussion in \cite{2006_07_MZ_DephasingNonGaussianNoise_NederMarquardt},
where the reader may find the relevant figures. 

At the outset, we note that the visibility in the real experiment
is also suppressed by external low-frequency fluctuations, beyond
the detector-induced dephasing discussed here. They contribute an
overall voltage-independent factor that has to be introduced as a
fitting parameter when comparing against theory. 

First, we consider the approximation (\ref{OneDetectingElectron})
obtained for low voltages, involving only one detector electron. Since
the constant $\gamma$ was measured, this formula does not contain
any free parameters, and can be compared directly with the experimental
data. As shown in Fig. 3 of \cite{2006_07_MZ_DephasingNonGaussianNoise_NederMarquardt},
it fits very well to the data at low $V$ and qualitatively reproduces
the novel effects mentioned at the beginning of section \ref{sec:Electronic-Mach-Zehnder-interferometer}.
In particular, it predicts the change from a smooth shape to a V-shape
in the dependence of visibility on partitioning probability, as well
as the non-monotonous behaviour with increasing $V$. However, according
to (\ref{OneDetectingElectron}) these effects should occur when $\gamma V=\pi$,
which does not agree with the experimental observations, where the
zero in the visibility is shifted to a detector bias that is larger
than this estimate by about 40\%. Hence at this detector bias (\ref{OneDetectingElectron})
fails to quantitatively reproduce the experimental results. The reason
of this discrepancy must be the onset of the contributions from other
detecting electrons. Once other eigenvalues become slightly non-zero,
the first one is smaller than $\gamma V$, because of the sum rule
(\ref{sumrule}) and the non-negativity of the eigenvalues, (\ref{nonnegative}).
This is clearly apparent in figure \ref{fig3}. The visibility then
vanishes at larger values of the detector voltage $V$, in agreement
with experiment. At even larger voltages, near $\varphi_{1}\approx2\pi$,
the visibility will have again a maximum (coherence revival). However,
it will be smaller due to the dephasing by the other detecting states
(other $\varphi_{j}$), again in contrast to the simplified formula
(\ref{OneDetectingElectron}). These two effects have both been seen
in the experiment (Fig. 4 in \cite{2006_07_MZ_DephasingNonGaussianNoise_NederMarquardt}). 

Finally, in \cite{2006_07_MZ_DephasingNonGaussianNoise_NederMarquardt},
we fitted the experimental data by using, for simplicity, a Lorentzian
shape as an ansatz for the Fourier transform of the phase function: 

\begin{equation}
\tilde{w}(q)=\frac{\tilde{w}(q=0)}{1+\left(\frac{qv_{{\rm det}}}{eV_{*}}\right)^{2}}\end{equation}
Here $V_{*}$ has the dimensions of a voltage and turns out to be
$V_{*}\approx6.2\,\mu V$. Within this fit, the first eigenvalue is
$\varphi_{1}\approx0.8\pi$ at $V=9.5\mu V$, where $\gamma V=\pi$.
This implies that almost the full phase shift of $\pi$ is contributed
by a single electron, indicating very strong interchannel interaction. \textbf{}

\subsection{Relation to intrinsic visibility oscillations}

\label{sub:Relation-to-intrinsic}While the earliest implementation
of the electronic MZI \cite{1998_02_Heiblum_WhichPath} displayed
a rather smooth monotonous decay of the visibility with rising MZI
bias voltage, this is no longer true in a more recent version \cite{2006_01_Neder_VisibilityOscillations}.
There, the visibility displayed oscillations, much like the ones observed
here, except they occured as a function of \emph{MZI bias voltage,}
in the \emph{absence} of any detector channel. The present analysis
may lead to a possible explanation for these initially puzzling observations:
The intrinsic \emph{intra-channel} interaction may cause the interfering
electrons to be dephased by their own (non-Gaussian) shot noise, if
the bias is large enough. We note that a similar explanation was put
forward in a recent preprint of Sukhorukov and Cheianov \cite{2006_09_Sukhorukov_MZ_CoupledEdges},
who considered a model where two \emph{counterpropagating} edge channels
interacted with each other. Though their model is therefore different
from ours, we have seen that the visibility oscillations are a generic
consequence of dephasing by non-Gaussian shot noise, and therefore
it is hard to distinguish experimentally (at this point) between the
different models.

\section{Summary and conclusions}

\label{sec:Summary-and-conclusions}We presented a nonperturbative
approach to the dephasing of a quantum system by an adjacent partitioned
one-dimensional electron channel, serving as a detector. Our treatment
gave an exact expression for the time-evolution of the visibility
of a charge qubit coupled to such a detector. Moreover, within a certain
simplifying approximation, it can be used to describe a {}``controlled
dephasing'' (or {}``which path'') setup where a Mach-Zehnder interferometer
is coupled to a detector channel. 

The main features of our results are the following: The visibility
may display oscillations as a function of time or detector voltage,
vanishing exactly at certain points and yielding {}``coherence revivals''
in-between those points. This behaviour is only observed if the coupling
strength crosses a certain voltage-independent threshold, corresponding
to a phase-shift of $g=\pi$ contributed by a single electron. It
is impossible to obtain that behaviour in any model of dephasing by
Gaussian noise, regardless of the assumed noise spectrum. The location
of the first zero of the visibility (in detector voltage or interaction
time) is proportional to $1/g$ for large couplings $g$, while the
spacing of subsequent zeroes is approximately independent of $g$
and corresponds to injecting one additional detector electron during
the interaction time. When plotted as a function of detector transmission
probability, the visibility differs from the smooth dependence on
$\mathcal{T}(1-\mathcal{T})$ expected for any Gaussian model, rather
displaying a {}``V-shape'' at certain voltages.

All of these features have been observed in the recent Mach-Zehnder
experiment \cite{2006_07_MZ_DephasingNonGaussianNoise_NederMarquardt}.
Challenges for future experiments include more quantitative comparisons
against the theory presented here, as well as finding ways of tuning
the interaction strength $g$, to switch between the strong and weak
coupling regimes. In addition, we hope that the strong coupling physics
of dephasing by non-Gaussian shot noise will be seen in future experiments
involving various other kinds of quantum systems as well.

\emph{Acknowledgements}. - We thank B. Abel, L. Glazman, D. Maslov,
M. B\"uttiker, Y. Levinson, D. Rohrlich, and M. Heiblum for fruitful
discussions. This work was partly supported by the Israeli Science
Foundation (ISF), the Minerva foundation, the German Israeli Foundation
(GIF), the SFB 631 of the DFG, and the German Israeli Project cooperation
(DIP).

\bibliographystyle{unsrt}
\bibliography{/Users/florian/pre/bib/shortall}

\end{document}